\documentclass[10pt]{article}
\usepackage{textcomp}
\usepackage{a4wide}
\usepackage{graphicx}
\usepackage{amsmath}
\usepackage{amsbsy}
\usepackage{epstopdf, epsfig}
\usepackage{amssymb}
\usepackage{longtable}
\usepackage{mathtools}
\usepackage{textcomp, csquotes}
\usepackage{xfrac} 
\usepackage{xcolor} 
\usepackage[inline]{enumitem}
\usepackage{afterpage}
\usepackage{natbib}

\usepackage{gensymb}
\usepackage{setspace}
\usepackage{geometry}
\usepackage{lineno}
\usepackage[hidelinks,colorlinks=true,linkcolor=blue]{hyperref}
\hypersetup{citecolor=blue, urlcolor=blue}
\usepackage{booktabs}

\usepackage{array}

\usepackage[caption=false]{subfig}
\definecolor{micro}{rgb}{0, 0, 0}
\definecolor{ure}{rgb}{0, 0, 0}
\definecolor{sub}{rgb}{0, 0, 0}
\definecolor{bio}{rgb}{0, 0, 0}
\definecolor{cal}{rgb}{0, 0, 0}
\definecolor{wat}{rgb}{0, 0, 0}
\definecolor{por}{rgb}{0, 0, 0}
\definecolor{per}{rgb}{0, 0, 0}
\definecolor{var}{rgb}{0, 0, 0}
\definecolor{co2}{rgb}{0, 0, 0}
\definecolor{alp}{rgb}{0, 0, 0}
\definecolor{david}{rgb}{1, .5, 0}

\newcommand{\co}{CO$_2  $}
\newcommand{\ctrlu}{x}

\newcommand{\ctrl}{\pmb{\ctrlu}}
\newcommand{\conctrl}{\hat{\ctrl}}
\newcommand{\conctrlu}{\hat{\ctrlu}}

\newcommand{\conctrlcomp}{\conctrl^{sol}}

\newcommand{\conctrlwat}{\conctrl^{wat}}

\newcommand{\conctrlnflw}{\conctrl^{nof}}
\newcommand{\nctrl}{N_{\ctrlu}}
\newcommand{\obj}{J}
\newcommand{\covu}{C}
\newcommand{\cov}{\pmb{\covu}}
\newcommand{\search}{\alpha}
\newcommand{\gradu}{g}
\newcommand{\grad}{\pmb{\gradu}}
\newcommand{\nens}{N_e}
\newcommand{\dt}{\Delta t}
\newcommand{\calvec}{\pmb{\phi}_c}

\newcommand{\calui}[1]{\phi_{c,#1}}
\newcommand{\nleak}{N_{l}}
\newcommand{\pen}{\gamma}
\newcommand{\pdmat}{\pmb{\Sigma}}
\newcommand{\nphase}{N_p}

\title{Field-scale optimization of injection strategies for leakage mitigation using microbially induced calcite precipitation}

\author{Svenn Tveit$^{1}$ \and and \and David Landa-Marb\'an$^{1}$}
\date{}
\begin{document}
\pagenumbering{arabic}
\maketitle
\onehalfspace
\normalsize
\noindent ${}^1$ NORCE Norwegian Research Centre AS, Nyg{\aa}rdsgaten 112, 5008 Bergen, Norway.\\[5pt]
Corresponding author: Svenn Tveit (E-mail address: svtv@norceresearch.no).

\begin{abstract}
\noindent An optimization procedure for sealing leakage paths in field-scale application of microbially induced calcite precipitation (MICP) is develop and applied to \co\ storage. The procedure utilizes a recently developed field-scale MICP mathematical model implemented in the industry-standard simulator Open Porous Media (OPM) Flow. The optimization problem is solved using the ensemble-based optimization (EnOpt) algorithm where the objective function is defined such that maximizing calcite precipitation is done in the shortest possible MICP operational time. An injection strategy is developed to efficiently produce calcite in and around the leakage paths, and to avoid clogging unwanted areas of the reservoir, e.g., the near-well area. The injection strategy consists of combined injection of growth and cementation solutions in separate well segments to initiate the MICP process after establishing a biofilm in the leakage paths with an initial injection phase. The optimization procedure is applied to three synthetic \co\ leakage scenarios. The numerical results show that the leakage paths in all scenarios are effectively sealed while keeping the total MICP operational time low. 
\end{abstract}

\section{Introduction}\label{Introduction}
To reduce the emission of anthropomorphic \co\ to the atmosphere, carbon capture and storage (CCS) in large, geological formations has been identified as a key remediation strategy \citep{Haszeldine2018}. The advantage of geological CCS is the huge volumetric storage potential in various saline aquifers or depleted oil and gas reservoirs. However, the effective storage potential of any geological formation depends, among other factors, on how much one can inject before reaching hazardous pressure build-ups. If supposedly closed leakage paths in the caprock, like faults, fractures, or abandoned wells, are exposed to pressures beyond their critical threshold, sequestrated \co\ might leak out of the storage site. Therefore, many in-depth case studies have been conducted on potential storage sites to ensure secure \co\ sequestration, e.g., \cite{Elenius2018},  \cite{Mulrooney2020}, and  \cite{Hodneland2019}. Even so, in the unlikely event that \co\ leakage paths may develop during injection, the consequences can be severe, both in terms of impact on the nearby environment and on the public acceptance of CCS. Thus, it is important to develop efficient and reliable leakage sealing technologies. 

\begin{figure}[h!]
	\centering
	\includegraphics[width=.2\textwidth]{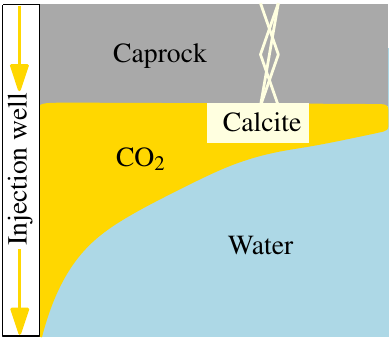}
	\caption{Visual representation of field-scale MICP application for \co\ leakage remediation}\label{fig:apl}
\end{figure}

A promising leakage sealing technology that has gained much attention in recent years is microbially induced calcite precipitation (MICP). The core idea of MICP is to use microbes to catalyse the chemical production of calcium carbonate -- calcite -- from urea and calcium to seal a leakage path \citep{Phillips2013}. Calcite is a low-permeable mineral that act as a sealing agent by reducing pore space, and as such, reduce the permeability of the leakage path. The technology has been proven effective both in column \citep{Cunningham2011} and core-scale studies \citep{Phillips2013a} for application in \co\ storage. Figure~\ref{fig:apl} shows a schematic representation of the MICP technology applied to CO$_2$ storage. It has also been used in several other applications, such as enhanced oil recovery \citep{Wu2017}, strengthening of liquefiable soil \citep{Burbank2011}, and concrete improvement \citep{DeMuynck2010}. Recently, \cite{Landa-Marban2021,Landa-Marban2021b} developed a numerical model to simulate the complex processes involved in field-scale application of MICP, based on work done in \cite{Ebigbo2012}, \cite{Hommel2015}, and \cite{Cunningham2019}. The model approximates the involved chemical and physical processes to capture the necessary field-scale behavior of MICP, enabling much shorter simulation times than previously developed pore- and core-scale models, e.g., \cite{Minto2019} and \cite{Nassar2018}.

\noindent Sealing leakage paths with MICP generally involves injecting several components into the reservoir: solution of pre-cultivated microbes; growth solution to establish biofilm (i.e., community of microbes) at leakage location; and cementation solution to initiate the MICP process in the biofilm. Several numerical studies have been conducted to develop injection strategies for MICP, e.g., \cite{Ebigbo2012}. The results from these studies have been applied in experiments from laboratory- to field-scale. However, most of the strategies focus on uniform sealing from the injection point, without completely plugging the inlet, e.g. for application of well fracture sealing \citep{Phillips2016}. In \co\ storage applications, leakage paths could develop tens-of-meters, or more, away from the injection well. Initiating the MICP process at the leakage location, without negatively impacting the rest of the storage site, is a challenging optimization task. Adding to the challenge is sealing the leakage paths in the least amount of time, to avoid a long shutdown of \co\ injection operation.

To perform optimization, standard gradient-based algorithms are often used. Even though many of the algorithms have good convergence properties, they need derivative calculations that are either computationally costly for field-scale models, or require access to simulation code for adjoint calculations. To address these issues, optimization methods with different stochastic approximations of the gradient have been developed. The advantages with most stochastic optimization methods are easy parallelization; they require only input-output interactions with a simulator; and multiple types of uncertainty, e.g., geological uncertainty, can be include in a straightforward manner. As a result, significant development have been made over the last decades, especially for the petroleum industry. In this paper, we apply the ensemble-based optimization (EnOpt) algorithm \citep{Chen2009} due to its simple implementation and it has been shown to perform well in benchmark studies, see, e.g., \cite{Chang2020}.

Few optimization studies involving MICP have been conducted in the literature. In \cite{Tveit2020} the authors investigated a risk-aware optimization workflow, where uncertainties on empirically determined model parameters could be included. In this paper, we focus on developing an optimization procedure for MICP with an injection strategy to avoid sealing unnecessary parts of the storage site. We apply the optimization procedure on synthetic test cases to gain insight on optimal injection strategies for field-scale applications of MICP. 

The paper is organized as follows: the different parts of the methodology is described in Section~\ref{Methodology}. This includes the mathematical model for MICP and its implementation in Sections~\ref{MICP mathematical model} and \ref{Implementation}, respectively. Section~\ref{Injection approach} describes the injection strategy, while a description of the different parts of the optimization method is given in Section~\ref{Optimization method}. The setup of the synthetic leakage scenarios for the numerical studies and subsequent results, together with a short discussion are given in Section~\ref{Results}. Lastly, we end the paper with some concluding remarks in Section~\ref{Conclusions}.

\section{Methodology}\label{Methodology}
We consider the optimization procedure for sealing leakage paths with the MICP technology. In the following sections, we describe the involved parts of the optimization procedure, that is, the field-scale simulation model for MICP processes, the injection strategy of the involved components, and the optimization method.    

\subsection{Mathematical model}\label{MICP mathematical model}

The conceptual model for field-scale application of the MICP processes is detailed in~\cite{Landa-Marban2021} and can be summed up as follows: microbial solution is injected into the reservoir to attach microbes to the rock at the leakage location; growth solution is injected to cultivate biofilm formation; and cementation solution is injected to initiate the calcite precipitation from the biofilm. The mathematical model only considers the rate-limiting components of the three solutions, which are suspended microbes, oxygen, and urea, respectively.

In this section, we give a brief overview of the mathematical model, and refer to \cite{Landa-Marban2021} for a complete description. The model is based on a preliminary study in \cite{Tveit2018}, together with previous MICP models in~\cite{Ebigbo2012}, \cite{Hommel2015}, and \cite{Cunningham2019}. Important additions to the model in \cite{Tveit2018} were dispersion, diffusion, and detachment by shear force. However, we recently observe from simulation studies that dispersive effects are more relevant for core-scale simulations, and less relevant at the field scale. Thus, we have removed the dispersion and diffusion terms in the mathematical model in this paper. Table \ref{tab1} shows all equations in the MICP mathematical model.

\begin{table*}[h!]
	\footnotesize
	\begin{center}
		\begin{minipage}{\textwidth}
			\caption{An overview of the equations in the MICP mathematical model\hspace{7cm}}\label{tab1}
			\begin{tabular*}{\textwidth}{@{\extracolsep{\fill}}m{2cm} m{12.7cm} m{.5cm}@{\extracolsep{\fill}}}
				\hline%
				Name & Equation &\\
				\hline
				Mass balance of water & $\partial_t {\color{por}\phi} +\nabla\cdot{\color{wat}\pmb{u}_w}=\underbrace{q_w}_{\text{Source term}},$\quad${\color{wat}\pmb{u}_w}=\underbrace{-\frac{{\color{per}\mathbb{K}}}{\mu_w}\left(\nabla {\color{wat}p_w}-\rho_w\pmb{g}\right)}_{\text{Darcy's law}}$ & \begin{equation}\label{eq:mass_bal}\end{equation}\\
				Mass balance of suspended microbes &  $\partial_t ({\color{micro}c_m}{\color{por}\phi})+\nabla\cdot({\color{micro}c_m}{\color{wat}\pmb{u}_w})=\underbrace{{\color{micro}c_m} q_w}_{\text{Source term}}+\underbrace{{\color{micro}c_m}{\color{por}\phi} Y\mu\frac{{\color{sub}c_o}}{k_o+{\color{sub}c_o}}}_{\text{Growth}}-\underbrace{{\color{micro}c_m}{\color{por}\phi} k_a}_{\text{Attachment}}+\underbrace{\rho_b{\color{bio}\phi_b} k_{str}({\color{por}\phi}\|\nabla {\color{wat}p_w}-\rho_w\pmb{g}\|)^{0.58}}_{\text{Detachment}}-\underbrace{{\color{micro}c_m}{\color{por}\phi} k_d}_{\text{Death}}$ & \begin{equation}\label{eq:micro}\end{equation}\\
				Mass balance of oxygen &    $\partial_t ({\color{sub}c_o}{\color{por}\phi})+\nabla\cdot({\color{sub}c_o}{\color{wat}\pmb{u}_w})=\underbrace{{\color{sub}c_o} q_w}_{\text{Source term}}-\underbrace{(\rho_b{\color{bio}\phi_b}+{\color{micro}c_m}{\color{por}\phi})F\mu\frac{{\color{sub}c_o}}{k_o+{\color{sub}c_o}}}_{\text{Consumption}}$  & \begin{equation}\end{equation}\\
				Mass balance of urea &  $\partial_t ({\color{ure}c_u}{\color{por}\phi})+\nabla\cdot({\color{ure}c_u}{\color{wat}\pmb{u}_w})=\underbrace{{\color{ure}c_u} q_w}_{\text{Source term}}\underbrace{-\rho_b{\color{bio}\phi_b}\mu_u\frac{{\color{ure}c_u}}{k_u+{\color{ure}c_u}}}_{\text{Urea conversion}}$  & \begin{equation}\end{equation}\\
				Mass balance of biofilm &  $\partial_t(\rho_b{\color{bio}\phi_b})=\underbrace{\rho_b{\color{bio}\phi_b}Y\mu\frac{{\color{sub}c_o}}{k_o+{\color{sub}c_o}}}_{\text{Growth}}+\underbrace{{\color{micro}c_m}{\color{por}\phi} k_a}_{\text{Attachment}}-\underbrace{\rho_b{\color{bio}\phi_b}k_{str}({\color{por}\phi}\|\nabla {\color{wat}p_w}-\rho_w\pmb{g}\|)^{0.58}}_{\text{Detachment}}-\underbrace{\rho_b{\color{bio}\phi_b}\left[k_d+\frac{{\color{cal}R_c}}{\rho_c(\phi_0-{\color{cal}\phi_c})}\right]}_{\text{Death}}$  & \begin{equation}\end{equation}\\
				Mass balance of calcite & $\partial_t(\rho_c{\color{cal}\phi_c})=\underbrace{\rho_b{\color{bio}\phi_b}Y_{uc}\mu_u\frac{{\color{ure}c_u}}{k_u+{\color{ure}c_u}}}_{\text{Produced calcite}}$  & \begin{equation}\label{eq:calcite}\end{equation}\\
				Porosity changes &  ${\color{por}\phi} =\underbrace{\phi_0-{\color{bio}\phi_b}-{\color{cal}\phi_c}}_{\text{Porosity reduction}}$  & \begin{equation}\label{eq:poro_change}\end{equation}\\
				Permeability changes & ${\color{per}\mathbb{K}}=\underbrace{\begin{cases}\left[\mathbb{K}_0\bigg(\frac{{\color{por}\phi} -\phi_{crit}}{\phi_0-\phi_{crit}}\bigg)^\eta+K_\text{min}\right]\frac{\mathbb{K}_0}{\mathbb{K}_0+K_\text{min}},&\phi_{\text{crit}}<{\color{por}\phi}\\ K_\text{min}, &{\color{por}\phi}\leq \phi_{\text{crit}}\end{cases}}_{\text{Porosity-permeability relationship}}$  & \begin{equation}\label{eq:perm_change}\end{equation}\\
				\hline
			\end{tabular*}
		\end{minipage}
	\end{center}
\end{table*}

In the mass balance equation of water, \eqref{eq:mass_bal}, the variables are the rock porosity $\phi$, the discharge per unit area $\pmb{u}_w$, and the source term $q_w$. $\pmb{u}_w$ is described by the Darcy's law where $\mathbb{K}$ is the rock permeability, $\mu_w$ the water viscosity, $p_w$ the water pressure, $\rho_w$ the water density, and $\pmb{g}$ the gravity vector. Furthermore, \eqref{eq:micro}--\eqref{eq:calcite} are mass balance equations for the suspended microbes (m), oxygen (o), urea (u), biofilm (f), and calcite (c), respectively. The notation for mass concentrations is $c_\xi$ ($\xi\in \lbrace m, o, u\rbrace$) and for volume fractions $\phi_\chi$ ($\chi\in \lbrace b, c\rbrace$). The right-hand side of these equations represent different phenomena during the MICP process, which are given under each term in Table \ref{tab1}. Here, $Y$ is the growth yield coefficient, $\mu$ the maximum specific growth rate, $k_o$ the half-velocity coefficient of oxygen, $k_d$ the microbial death coefficient, $k_a$ the microbial attachment coefficient, $k_{str}$ the detachment rate, $F$ the mass ratio of oxygen consumed to substrate used for growth, $\mu_u$ the maximum rate of urea utilization, $k_u$ the half-velocity coefficient for urea, and $Y_{uc}$ the yield coefficient for the produced calcite over the urea utilization. Lastly, \eqref{eq:poro_change} represents the porosity change due to increasing biofilm and calcite volume fractions, and \eqref{eq:perm_change} models the change in permeability due to changes in porosity, where  $\phi_{crit}$ is the critical porosity
when the permeability becomes a minimum value $K_{min}$. In Table \ref{tab2} values for all model parameters are given. Full references to the sources for these values can be found in \cite{Landa-Marban2021}.

\begin{table*}[h!]
	\begin{minipage}{\textwidth}
		\caption{Model parameters for the mathematical model in Table~\ref{tab1}\hspace{8cm}}\label{tab2}%
		\begin{tabular}{@{} m{5cm} m{1cm} m{2cm} m{1cm} @{}}
			\hline
			Parameter & Sym. & Value & Unit\\
			\hline
			Density (biofilm) &$\rho_b$ & $35$ & $\sfrac{\text{kg}}{\text{m}^3}$\\
			Density (calcite)& $\rho_c$	& $2710$ & $\sfrac{\text{kg}}{\text{m}^3}$\\
			Density (water) &$\rho_w$	& 	$1045$ & $\sfrac{\text{kg}}{\text{m}^3}$\\
			Detachment rate & $k_{str}$ & 2.6$\times10^{-10}$ & $\sfrac{\text{m}}{\text{Pa s}}$\\
			Critical porosity & $\phi_\text{crit}$ & 0.1& [-]\\
			Fitting factor & $\eta$ & 3& [-]\\
			Half-velocity coefficient (oxygen)&$k_o$	& 	$2\times 10^{-5}$ & $\sfrac{\text{kg}}{\text{m}^3}$\\
			Half-velocity coefficient (urea)& $k_u$	&	$21.3$ & $\sfrac{\text{kg}}{\text{m}^3}$\\
			Maximum specific growth rate &$\mu$ &    $4.17\times 10^{-5}$& $\sfrac{1}{\text{s}}$\\
			Maximum rate of urea utilization &$\mu_u$ &    $1.61\times 10^{-2}$ & $\sfrac{1}{\text{s}}$\\
			Microbial attachment rate &$k_a$ &    $8.51\times 10^{-7}$ & $\sfrac{1}{\text{s}}$\\
			Microbial death rate&$k_{d}$ & 	$3.18\times 10^{-7}$ & $\sfrac{1}{\text{s}}$\\ 
			Minimum permeability& $K_\text{min}$	& $10^{-20}$ & $\text{m}^2$\\
			Oxygen consumption factor&$F$ & 	$0.5$ &  [-]\\ 
			Water viscosity &$\mu_w$	& 	$2.54\times10^{-3}$ & $\textrm{Pa s}$\\
			Yield coefficient (growth) &$Y$	& 	$0.5 $ & [-]\\
			Yield coefficient (calcite/urea) &$Y_{uc}$	& 	$1.67 $ & [-]\\
			\hline
		\end{tabular}
	\end{minipage}
\end{table*}

\subsection{Implementation}\label{Implementation}
The mathematical model in Section~\ref{MICP mathematical model} was implemented in the industry-standard, open-source simulator Open Porous Media (OPM) Flow. The simulator is part of the OPM initiative, that facilitates field-scale simulations of different subsurface applications such as hydrocarbon recovery and geological CO$_2$ storage. It is a fully-implicit, finite difference simulator on corner-point grids, with advanced well models. We refer to \cite{Rasmussen2019} for a description of OPM Flow along with some of the implemented models, such as the black-oil and well models. 

\begin{figure}[h!]
	\centering
	\includegraphics[width=0.6\textwidth]{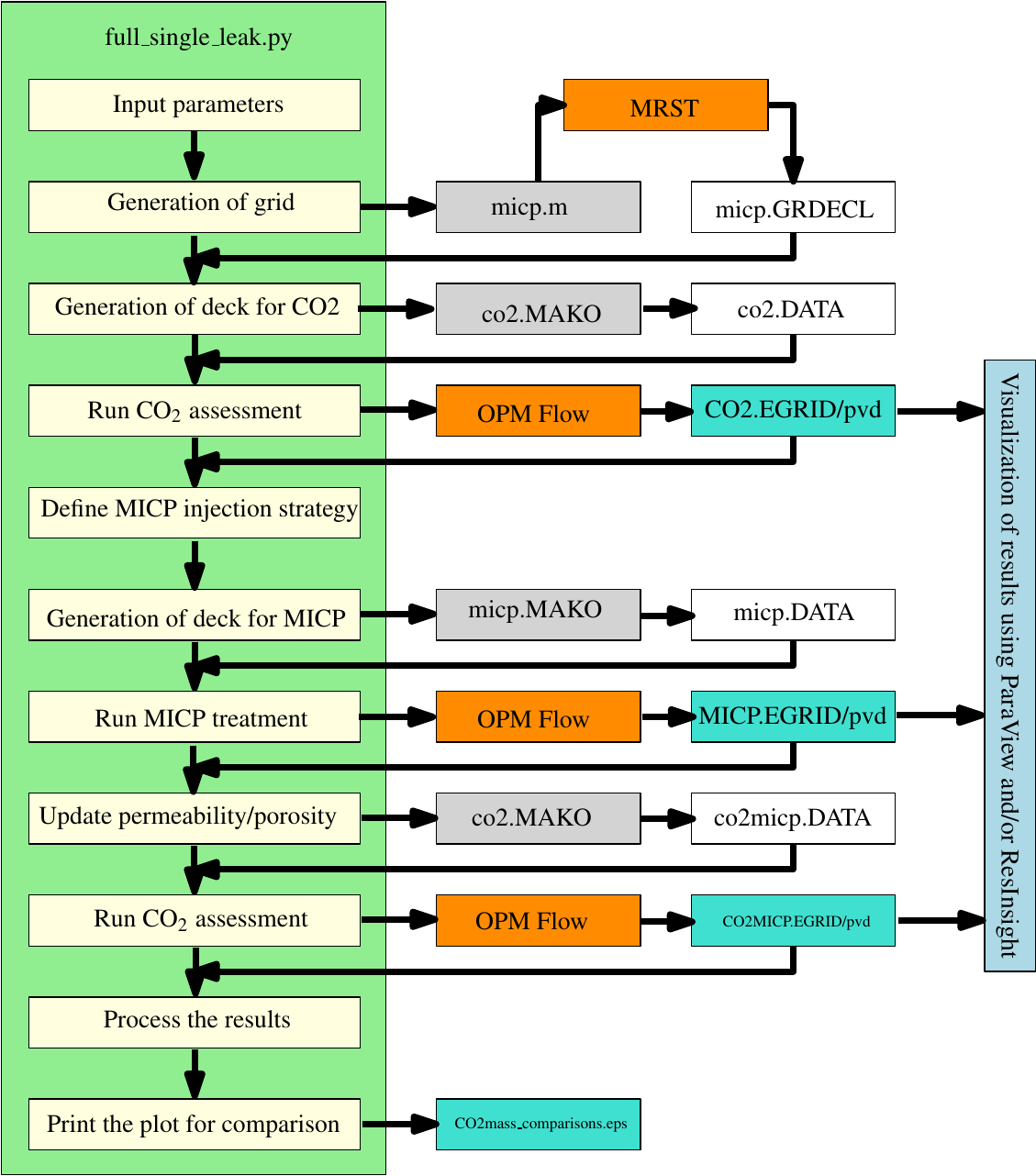}
	\caption{Workflow example of \texttt{py-micp}}\label{figfra}
\end{figure}

The mathematical model for MICP described in Section~\ref{MICP mathematical model} was made available in the 2021.10 release of OPM Flow. The model in \cite{Landa-Marban2021} was made available in the 2021b release of the Matlab Reservoir Simulation Toolbox (MRST). In \cite{Landa-Marban2021b} we show comparison between simulations in MRST and OPM for a simple 1D horizontal system, resulting in a good agreement between numerical results from both simulators. A description of the OPM Flow keywords for the MICP model can be found in the OPM Flow manual \citep{Baxendale2021}. Lastly, we have developed a Python package, \texttt{py-micp}, to integrate different open-source code to perform studies of MICP treatment and CO$_2$ assessment. Figure \ref{figfra} shows an example of the workflow on a case study.

\subsection{Injection strategy}\label{Injection approach}
In \cite{Landa-Marban2021}, the authors developed a field-scale injection strategy of the microbial, growth, and cementation solutions, based on earlier experimental and numerical studies. The strategy developed there involved injecting each solution separately, with follow-up injection of water and no-flow periods. The numerical studies performed in \cite{Landa-Marban2021}, showed that even leakage paths considerably far from the injection well were completely sealed with the proposed strategy. Furthermore, it was shown that injecting solutions only in the top segment of the injection well, and water in the rest of the segments, was beneficial to avoid calcite precipitation along the whole vertical direction of the reservoir.

In this paper, we further develop the injection strategy introduced in \cite{Landa-Marban2021}. Let a phase be defined as an injection of one or more solutions with subsequent water displacement and no-flow periods. Furthermore, let $ t_1^p $ be the end time of solution injection, $ t_2^p $ be the end time of water displacement, and $ t_3^p $ be the end time of the no-flow period for phase $ p $. Lastly, we assume that the solutions can be injected in different segments of the injection well, and that the injection rate is constant. 

\begin{figure*}[h!]
\centering
\includegraphics[width=\textwidth]{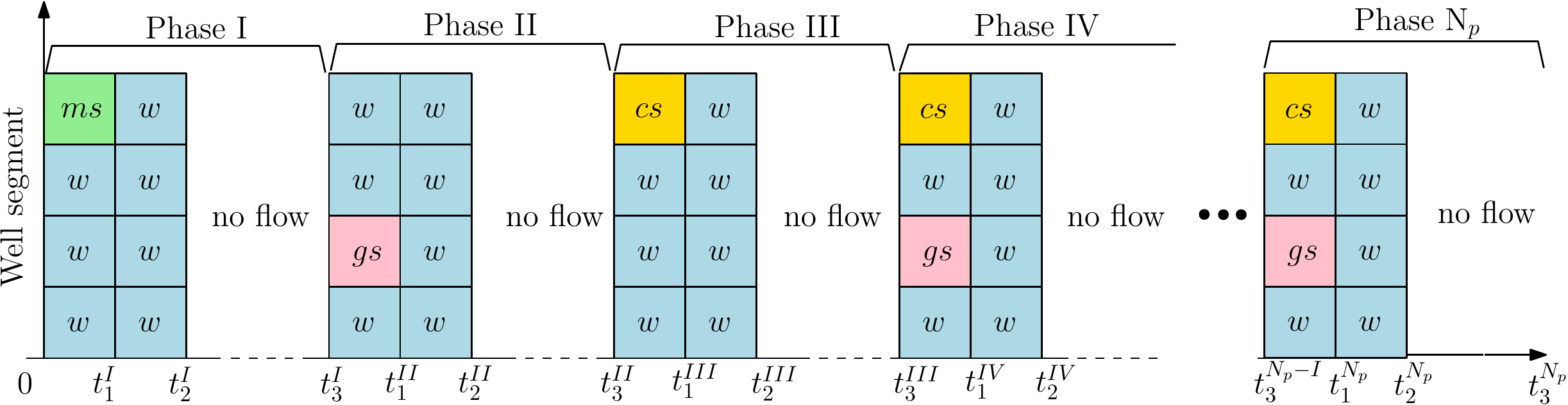}
\caption{An illustration of the injection strategy where ms, gs, cs, and w refer to injection of microbial, growth, and cementation solutions, and only water, respectively}\label{fig:ins}
\end{figure*}

The injection strategy proposed in this paper is then defined in Figure~\ref{fig:ins}. From the figure we see that phases $ I $--$ III $ involve separate injection of microbial, growth, and cementation solutions, respectively. This follows the strategy in \cite{Landa-Marban2021}, with the goal of developing biofilm and subsequently produce calcite at the leakage paths. In phases $ IV $--$ \nphase $ we inject growth and cementation solutions at the same time, to utilize the remaining microbes and biofilm at the leakage paths to produce calcite. Moreover, the growth solution is injected in a lower segment of the injection well than the microbial and cementation solutions. Since the permeability of the leakage paths are higher than the rest of the reservoir, the growth solution will be transported to the leakage location under correct flow conditions. As a result, we limit the MICP processes to only occur in and around the leakage paths, instead of potentially initiating the process in unwanted areas, if all solutions are injected in the same well segment.

\subsection{Optimization method}\label{Optimization method}
Let $ \obj(\ctrl) $ denote the scalar objective function, with $ \ctrl \in \mathbb{R}^{\nctrl}$ being the control variables. The optimization problem is then to maximize $ \obj(\ctrl) $ with respect to $ \ctrl $. A widely-used method for optimization, with guarantied local convergence, is the pre-conditioned steepest ascent method,

\begin{equation}\label{eq:steep_asc}
\ctrl_{k+1} = \ctrl_{k} + \search_k\pdmat\grad_k,
\end{equation}

where $ k $ is the iteration number, $ \search_k $ is the step size,  $ \pdmat $ is a symmetric, positive definite matrix, and $ \grad_k $ is the gradient of $ \obj(\ctrl_k) $. To solve \eqref{eq:steep_asc}, EnOpt \citep{Chen2009} approximates $ \pdmat\grad_k $ by the sample cross-covariance matrix. In this paper, we use the modified version from \cite{Do2013} given by

\begin{equation}\label{eq:enopt}
\pdmat\grad_k \approx \cov_{\ctrlu, \obj} = \frac{1}{\nens}\sum_{j=1}^{\nens} \left(\ctrl_{k}^j - \ctrl_k\right) \left(\obj(\ctrl_k^j) - \obj(\ctrl_{k}) \right).
\end{equation}

Here, $ \{\ctrl_k^j\}_{j=1}^{\nens} $ is sampled from a multivariate normal distribution with mean $ \ctrl_k $ and covariance matrix $ \pdmat $. To ensure that the approximation in \eqref{eq:enopt} is reasonable, sufficiently large $ \nens $ and small sample perturbations when generating the $ \ctrl_k^j $'s are required \citep{Do2013}. In practice, $ \nens $ is chosen based on the available computational budget, since a minimum of $ \nens + 1$ evaluations of $ \obj(\ctrl) $ is needed per update in \eqref{eq:steep_asc} with \eqref{eq:enopt}. For a thorough theoretical discussion on EnOpt see \cite{Stordal2016}.

To enforce upper and lower bounds on the control variables, we use the log-transform defined in \cite{Do2013},

\begin{equation}\label{eq:log}
\ctrlu_i = \log\left(\frac{\conctrlu_i - \conctrlu_i^{low}}{\conctrlu_i^{up} - \conctrlu_i}\right),
\end{equation}

where $ \conctrlu_i $ denotes the constrained control variable, and $ \conctrlu_i^{up}$ and $ \conctrlu_i^{low} $ are the upper and lower bounds, respectively. Hence, the optimization using $ \ctrl $ will be unconstrained since $ \ctrlu_i \rightarrow -\infty $ when $ \conctrlu_i \rightarrow \conctrlu_i^{low}$, and $ \ctrlu_i \rightarrow \infty $ when $ \conctrlu_i \rightarrow \conctrlu_i^{up}$. To invert $ \ctrl $ to $ \conctrl $, e.g., in evaluations of $ \obj(\ctrl) $, we can use the inverse formula to \eqref{eq:log},

\begin{equation}\label{eq:inv_log}
\conctrlu_i = \frac{\exp(\ctrlu_i)\conctrlu_i^{up} + \conctrlu_i^{low}}{1 + \exp(\ctrlu_i)}.
\end{equation}

A widely-used application of optimization on field scale is production from petroleum reservoirs. Here, the objective function $ \obj(\ctrl) $ is typically defined through an economic model given by the net present value (NPV). This model has been expanded with additional terms depending on new applications, e.g., with costs and revenues related to \co\ sequestration \citep{Chen2020}. However, to optimize the sealing of leakage paths with MICP, a pure economic model may not be beneficial, since such operations are merely cost. We suggest instead combining a physical property model for calcite precipitation with a penalty term on the total number of MICP operational days in the definition of $ \obj(\ctrl) $. Hence, the main aim of the optimization is to maximize the sealing of leakage paths, but at the same time minimize the injection and no-flow periods. With the penalty term we add an economic aspect to our optimization, since reducing operational days of MICP remediation will reduce the cost.

To this end, let $ \conctrl $ be the injection and no-flow periods defined in Section \ref{Injection approach}. For ease of reading, we divide $ \conctrl $ in three parts $ \conctrl = [\conctrlcomp, \conctrlwat, \conctrlnflw] $ and define each one as follows: $ \conctrlcomp = [\dt^I_{1,3}, \dt^{II}_{1,3}, \ldots, \dt^{\nphase}_{1,3}] $ where $ \dt^p_{1,3} = t^p_1 - t^{p-I}_3 $ with $ t^0_3 = 0 $; $ \conctrlwat = [\dt_{2,1}^I, \dt_{2,1}^{II}, \ldots, \dt_{2,1}^{\nphase}] $ where $ \dt_{2,1}^p = t_2^p - t_1^p $; and $ \conctrlnflw = [\dt_{3,2}^I, \dt_{3,2}^{II}, \ldots, \dt_{3,2}^{\nphase}], $ where $ \dt_{3,2}^p = t_3^p - t_2^p $. In short, $ \conctrlcomp $ are the injection periods for the microbial, growth, and cementation solutions; $ \conctrlwat $ are the water displacement periods; and $ \conctrlnflw $ are the no-flow periods. Furthermore, let $ \calvec^{leak}(t_3^p) = [\calui{1}^{leak}(t_3^p), \calui{2}^{leak}(t_3^p), \ldots, \calui{\nleak}^{leak}(t_3^p)]$ where $ \calui{i}^{leak}(t_3^p) $ is the calcite volume fraction in cell $ i $ inside the leakage path at time $ t_3^p $. The objective function is then defined as

\begin{equation}\label{eq:obj}
\obj(\ctrl) = \sum_{p} \|\calvec^{leak}(t_3^p)\|_{\infty} - \pen\|\conctrl\|_1,
\end{equation}

where $ \pen $ is a constant weighting parameter. Hence, the first term in \eqref{eq:obj} is the sum of the maximum calcite volume fraction in the leakage paths at the end of a phase (indicated by $ t_3^p $). Note that from \eqref{eq:perm_change} the maximum value for $ \calui{i}^{leak}(t_3^p)$ is $ \phi_0 - \phi_{crit} $.

\section{Numerical experiments}\label{Results}
\begin{figure}[h!]
	\centering
	\includegraphics[width=.45\textwidth]{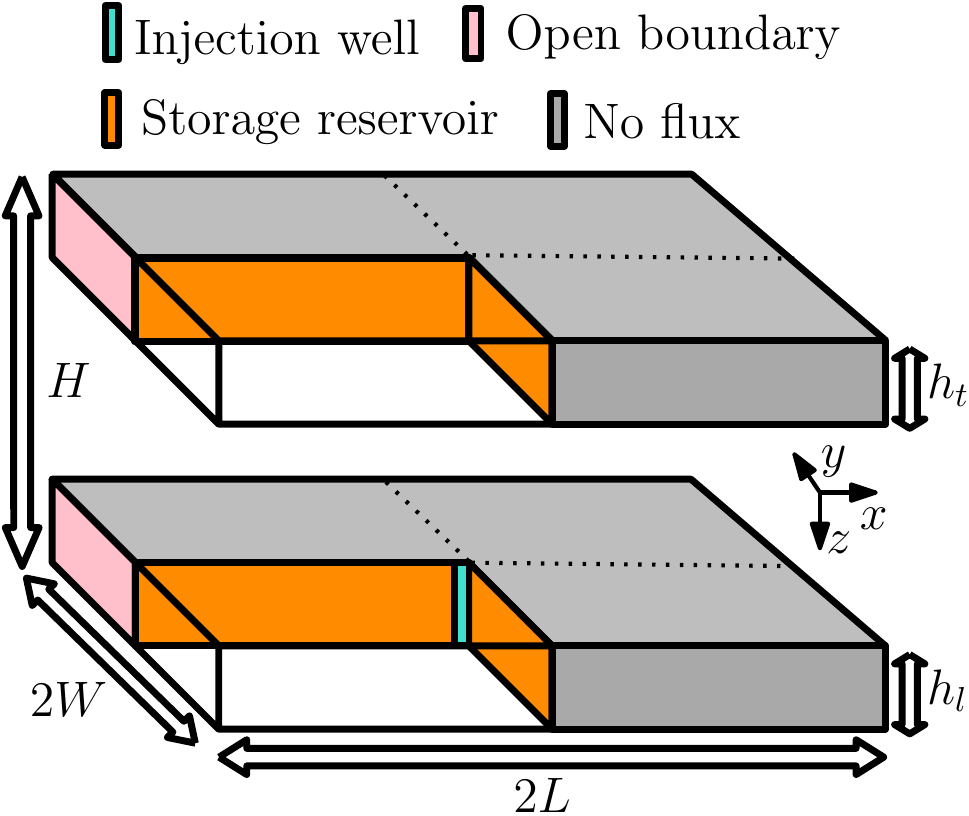}
	\caption{Illustration of the full 3D system with two aquifers separated by a caprock (not shown explicitly). The quarter domain considered in the numerical experiments is outlined by the dashed lines}\label{fig:reservoir}
\end{figure}

In this section, we apply the optimization procedure to seal leakage paths in three different scenarios. A synthetic reservoir was set up similar to the 3D system studied in \cite{Landa-Marban2021}, which was based on the \co\ benchmark study in \cite{Ebigbo2007} and \cite{Class2009}. The reservoir consisted of two aquifers separated by a caprock, where we in this study had three scenarios with different leakage paths through the caprock. To reduce the computational time, we only considered a quarter of the full system; a common approximation in the literature, see, e.g., \cite{Zhang2012}. Examples considering the full system can be found in \texttt{py-micp}, and they confirm that the quarter-system approximation is valid in the three leakage scenarios. Thus, the computational domain was a 37$ \times $37$ \times $30 grid, with equidistant 1m$ \times $1m$ \times $1m cells close to the injection wells and exponentially increasing cell sizes in $ x$- and $y $-directions towards the boundaries. Note that the caprock is modeled with non-active cells, thus the computational domain consisted of only 13~710 active cells. The injection well is located in the lower left corner of the domain, which had a constant rate of $ 2\cdot10^{-2} $~m/s during the injection periods, and a production well was placed in the upper, right corner to simulate open boundaries in OPM Flow, since it has fixed no-flow boundary conditions. A summary of the reservoir and fluid properties is given in Table~\ref{tab:res_prop}, and an illustration of the full system is shown in Figure~\ref{fig:reservoir}, where the upper right quarter was used in the scenarios. Note that we assume constant permeability in each cell, thus the tensor $ \mathbb{K}_0 $ in \eqref{eq:perm_change} reduces to a scalar $ K_0 $.

\begin{table}[h!]
	\begin{minipage}{\columnwidth}
		\caption{Reservoir and fluid properties\hspace{12cm}}\label{tab:res_prop}
		\begin{tabular}{@{} m{4.5cm} m{1cm} m{1.5cm} m{1.0cm} @{}}
			\hline
			Parameter & Sym. & Value & Unit\\
			\hline
			Height domain &$H$ & $30$ & m\\
			Height aquifers &$h_l/h_t$ & $5$ & m\\
			Width/length &$W/L$ & $100$ & m\\
			Initial permeability aquifer& $K_0^A$	& $1\cdot10^{-14}$ & $\text{m}^2$\\
			Initial permeability leakage& $K_0^L$	& $2\cdot10^{-14}$ & $\text{m}^2$\\
			Initial porosity & $\phi_0$ & 0.15& [-]\\
			Microbial concentration & $ c_m $ & 0.01 & $ \text{kg}/\text{m}^3 $ \\
			Oxygen concentration & $ c_o $ & 0.04 & $ \text{kg}/\text{m}^3 $ \\
			Urea concentration & $ c_u $ & 60.0 & $ \text{kg}/\text{m}^3 $ \\
			\hline
		\end{tabular}
	\end{minipage}
\end{table}

\begin{figure*}[h!]
	\subfloat[]{\includegraphics[width=0.33\textwidth]{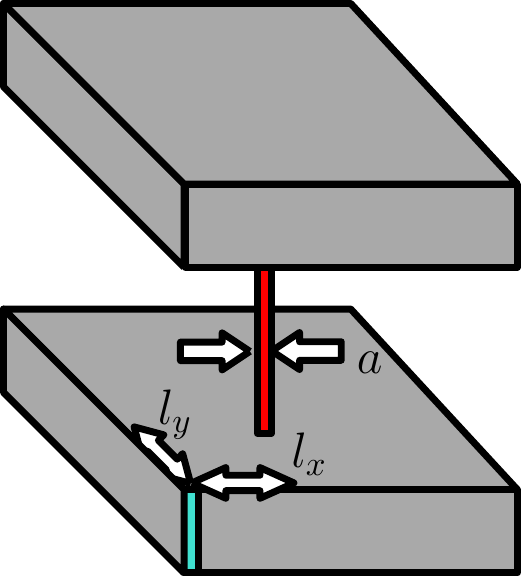}}
	\subfloat[]{\includegraphics[width=0.33\textwidth]{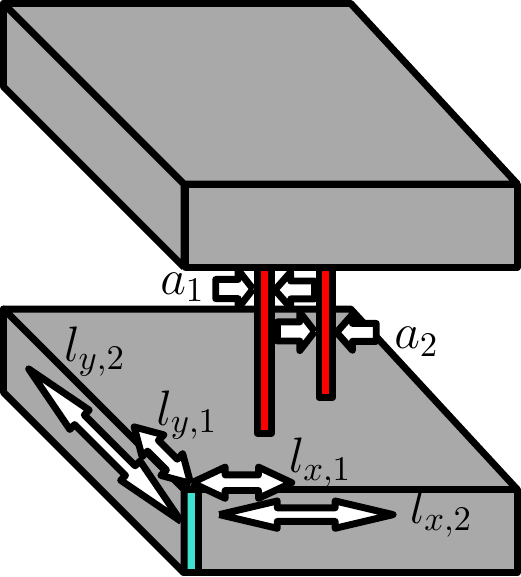}}
	\subfloat[]{\includegraphics[width=0.33\textwidth]{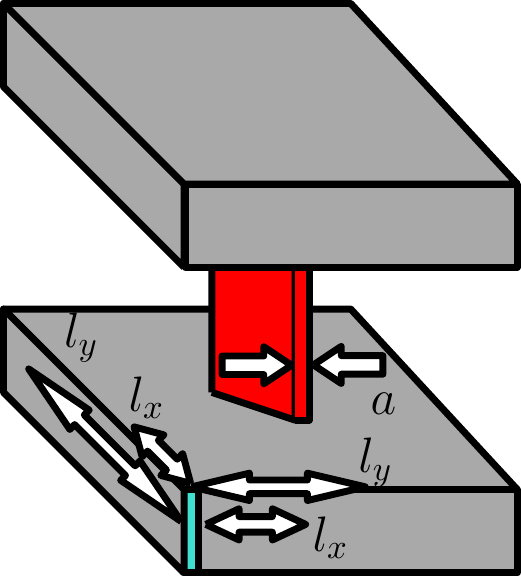}}
	\caption{Illustrations of the 3D domains for the (a) Single, (a) Double, and (c) Diagonal leak scenarios}\label{fig:three_scenarios}
\end{figure*}

The three scenarios were set up as follows: (i) a single leakage path, similar to the leaky well scenario considered in \cite{Landa-Marban2021}; (ii) two separate leakage paths; and (iii) a wide, connected leakage path, similar to the one considered in \cite{Landa-Marban2021b}. For convenience, we label scenarios (i), (ii), and (iii) as Single leak, Double leak, and Diagonal leak, respectively. In Figure~\ref{fig:three_scenarios}, illustrations of the three scenarios are shown, with properties for each scenario summarized in Table~\ref{tab:scenario_prop}. Due to computational constraints, we considered leakage paths that are relatively large in size, especially compared to typical fractures sizes. Thus, the scenarios can be seen as mimicking three types of damaged zones in the caprock, with a slightly higher permeability ($ K_0^L = 2\cdot K_0^A $; see Table~\ref{tab:res_prop}).

\begin{table*}[h!]
	\begin{minipage}{\textwidth}
		\caption{Domain properties for the three leakage scenarios. All values are in meters\hspace{5cm}}\label{tab:scenario_prop}
		\begin{tabular*}{\textwidth}{@{\extracolsep{\fill}} l l l l l l l l l l l l l l @{\extracolsep{\fill}}}
			\hline
			\multicolumn{3}{@{}c@{}}{Single leak} & & \multicolumn{6}{@{}c@{}}{Double leak} & & \multicolumn{3}{@{}c@{}}{Diagonal leak} \\
			\cmidrule{1-3} \cmidrule{4-10} \cmidrule{11-14}
			$a$ & $l_{x}$ & $l_{y}$ & & $a_1$ & $a_2$ & $l_{x,1}$ & $l_{y,1}$ & $l_{x,2}$ & $l_{y,2}$ & & $a$ & $l_{x}$ & $l_{y}$ \\
			\cmidrule{1-3} \cmidrule{4-10} \cmidrule{11-14}
			1 & 13 & 14 & & 1 & 1 & 10 & 11 & 17 & 18 & & 1 & 11 & 19 \\
			\hline
		\end{tabular*}
	\end{minipage}
\end{table*}

\noindent We applied the injection strategy described in Section~\ref{Injection approach} with $ \nphase = 5 $. Hence, we followed up phases $ I $--$ III $ with two phases, $ IV $ and $ V $, of combined injection of growth and cementation solutions. Note that more follow-up phases to $ I $--$ III $ could be considered in some applications to ensure complete sealing of leakage paths. Examples of ad-hoc injection strategies using additional phases to fully seal the leakage paths in the three scenarios can be found in \texttt{py-micp}. However, that could also lead to a lot of wasted solution if there are no microbes and/or biofilm at the leakage locations. Furthermore, the extended MICP operation time with more phases can lead to little benefit in terms of further sealing the leakage paths, and thus be unsound from an economical perspective.

In all scenarios, the control variables for phases $ I $--$ V $, i.e., the injection and no-flow periods in each phase, were constrained with lower bounds all equal to 0~d, and upper bounds equal to 5~d for $ \conctrlcomp $ and $ \conctrlwat $, and 6~d for $ \conctrlnflw $. For the objective function, $ \obj(\ctrl) $, we define the cells in $ \calvec^{leak} $ as the first cells in the leakage paths immediately above the lower aquifer. Hence, we want to seal the leakage paths efficiently by maximizing the calcite precipitation in and around the entry of the leakage paths. We evaluated $ \calvec^{leak} (t_3^p) $ at $ p =  III, IV, V$. Furthermore, $ \pen $ was chosen in all scenarios to weight the first term in \eqref{eq:obj} higher than the second term. In preliminary studies, we have seen that too much weight on the second term may lead to optimization results with little or no calcite precipitation due to very short injection and no-flow periods.

EnOpt was set up with $ \nens = 20 $ and $ \pdmat = \sigma\pmb{I} $, where $ \sigma=0.01 $. The optimization procedure was terminated when one or more of the following criteria were reached: 

\begin{enumerate}
	\item An uphill direction (i.e, $ \obj(\ctrl_{k+1}) > \obj(\ctrl_{k})$), could not be found with at least ten cuts to the step size, $ \search_k $
	\item $ \lvert \obj(\ctrl_{k+1}) - \obj(\ctrl_{k}) \rvert / \lvert \obj(\ctrl_{k}) \rvert \leq 10^{-6}$
	\item $\|\ctrl_{k+1} - \ctrl_{k}\|_2 \leq 10^{-6} $
	\item Total number of iterations, including step-size cuts, reaches 50
\end{enumerate}

To assess the quality of the optimization results, we ran simulations of \co\ injection in the reservoir before and after MICP using the final control variables from the optimization. We injected \co\ for 400~d with injection rate equal to $ 1\cdot10^{-4} $~m$ ^3 $/s. To quantify the reduction in leakage of \co\ through the leakage paths before and after MICP, we calculated the percentage reduction in accumulated \co\ mass in the upper aquifer, given in Table~\ref{tab:opt_results}.

\begin{table*}[h!]
	\begin{center}
		\begin{minipage}{\textwidth}
			\caption{Optimization results for all scenarios. For convenience, the injection and no-flow periods in $ \ctrl $ have been converted to times $ t_1$, $t_2 $, and $ t_3 $ for each phase in hours; see Section~\ref{Injection approach}}\label{tab:opt_results}
			\begin{tabular*}{\textwidth}{@{\extracolsep{\fill}}l l r r r c c@{\extracolsep{\fill}}}
				\hline
				Scenario & Phase & \multicolumn{3}{@{}c@{}}{Times [h]} &   \multicolumn{2}{@{}c@{}}{Leaked \co\  [\%]} \\
				\cmidrule{3-5} \cmidrule{6-7}
				&  & $t_1^p\;$& $t_2^p\;$& $t_3^p$ & before & after\\
				\hline
				& $I$ & 14.93 & 15.35 & 17.87 &    \\
				& $II$ & 33.24 & 34.97 & 36.38 &    \\
				Single leak & $III$ &  36.78 & 37.91 & 43.92 &  4.30 & 6.67$\cdot10^{-12}$\\
				& $IV$ &  155.11 & 156.04 & 157.88 &    \\
				& $V$ &  159.11 & 160.27 & 161.52 &    \\
				\\
				& $I$ & 13.70 & 14.16 & 22.84 &    \\
				& $II$ & 53.87 & 54.48 & 71.28 &    \\
				Double leak & $III$ &  85.25 & 87.18 & 102.84 &  5.29 & 2.76$\cdot10^{-10}$ \\
				& $IV$ &  183.11 & 184.12 & 191.44 &    \\
				& $V$ &  196.33 & 198.83 & 207.65 &    \\
				\\
				& $I$ & 4.97 & 6.83 & 23.70 &    \\
				& $II$ & 58.30 & 59.01 & 74.58 &    \\
				Diagonal leak & $III$ &  83.92 & 85.25 & 86.74 &  7.04 & 1.28$\cdot10^{-9}$ \\
				& $IV$ &  191.26 & 191.83 & 194.23 &    \\
				& $V$ &  195.40 & 196.98 & 199.50 &    \\
				\hline
			\end{tabular*}
		\end{minipage}
	\end{center}
\end{table*}

\subsection{Single leak}\label{Scenario i}
\begin{figure}[h!]
\centering
	\subfloat[\label{fig:cs_single}]{\includegraphics[width=0.33\textwidth]{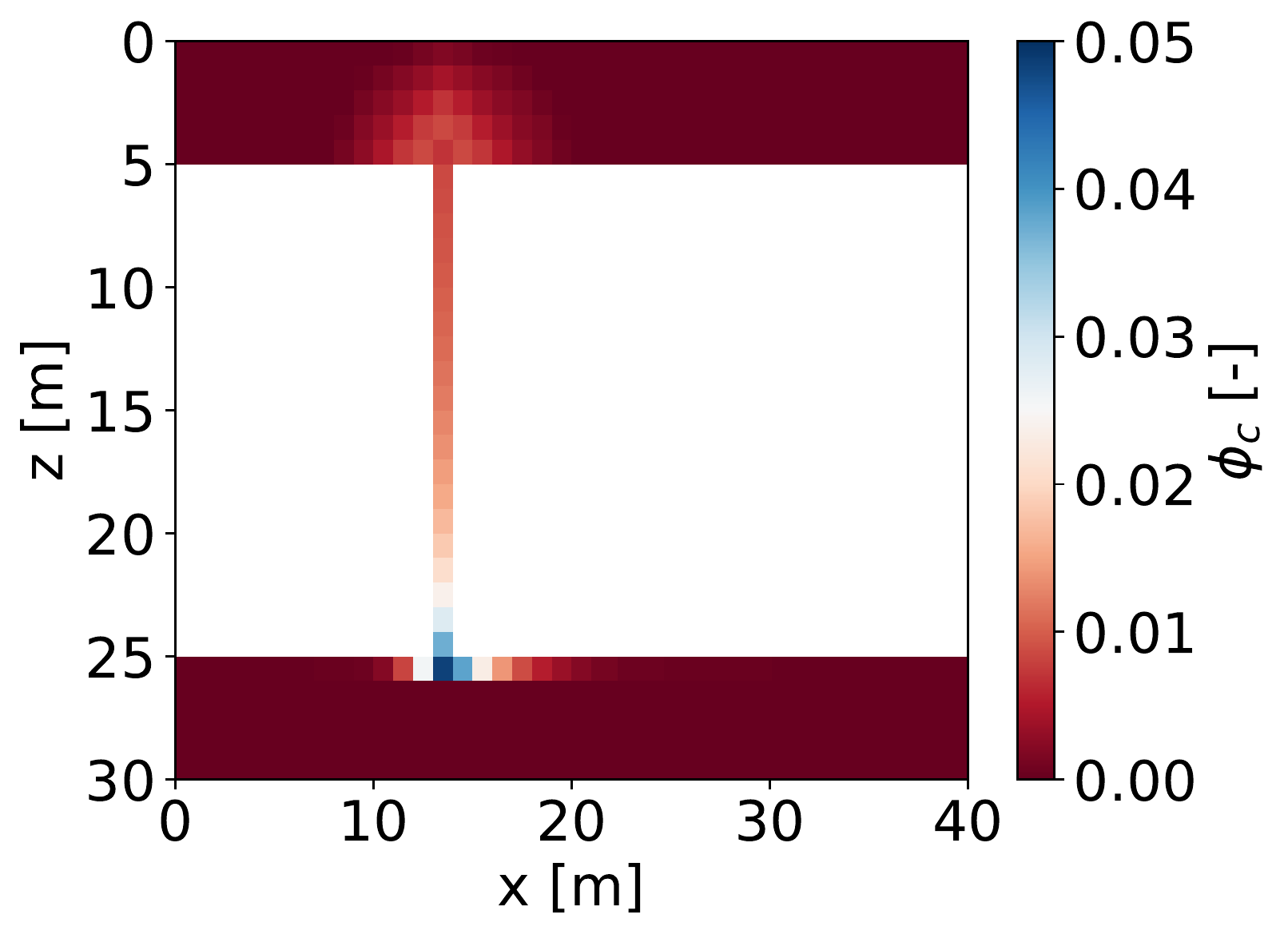}}\quad    
	\subfloat[\label{fig:obj_single}]{\includegraphics[width=0.335\textwidth]{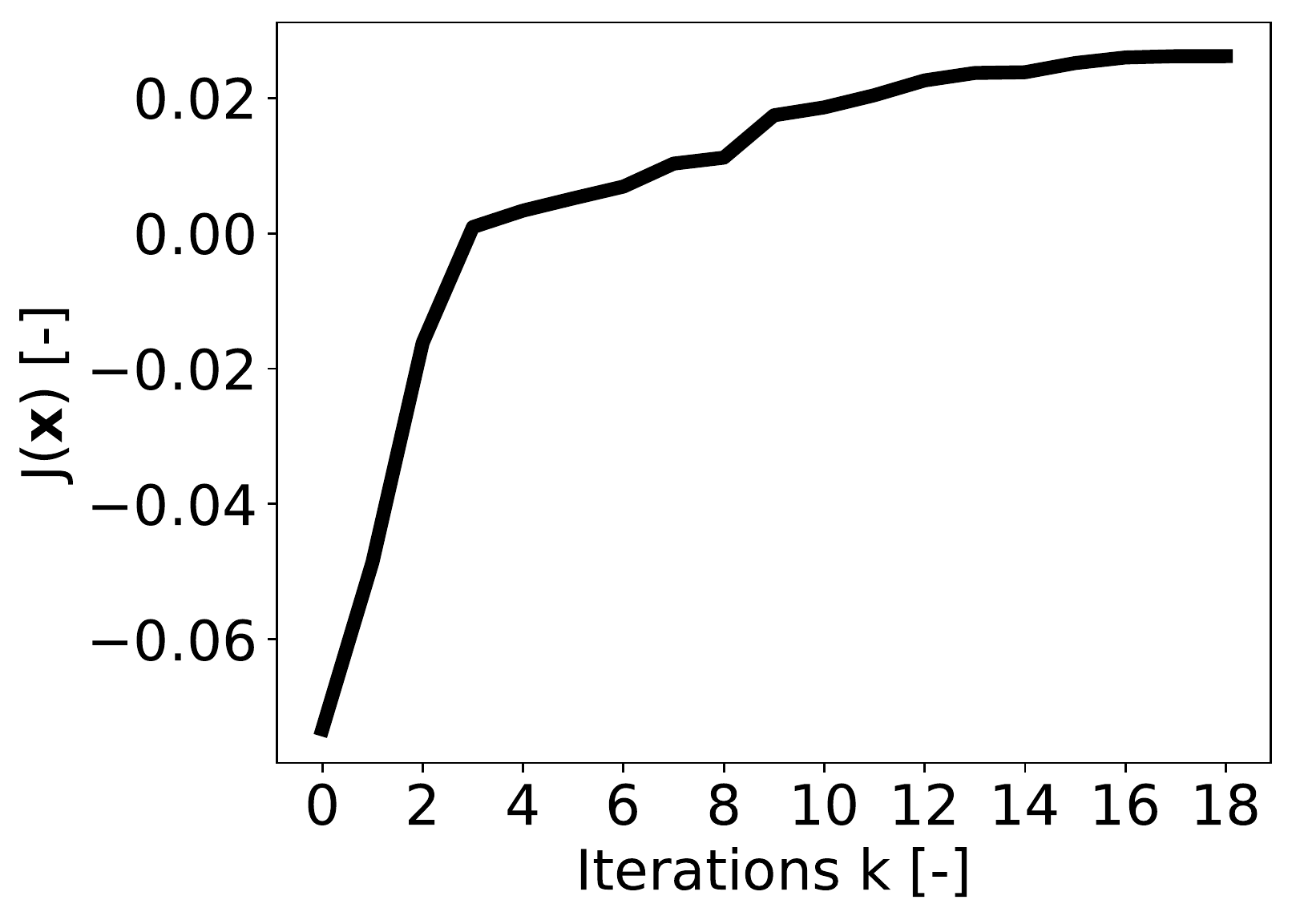}}
	\caption{Single leak: (a) vertical cross-section plot at y = 14 m of the calcite distribution using the final optimization results and (b) objective function values versus iterations}
\end{figure}

In this scenario, the initial control variables were 1~d for $ \conctrlcomp_0 $, 0.05~d for $ \conctrlwat_0 $, and 2~d for $ \conctrlnflw_0 $, and $ \pen = 7.14\cdot10^{-3} $~d$^{-1}$. The final optimization results, converted to times $ t_1^p $, $ t_2^p $, and $ t_3^p $, are shown in Table~\ref{tab:opt_results}. We see that the total time, $ t_3^{V} $, is 6.73~d (or 161.52~h), which is a significant reduction from the total time of the initial control variables, 15.25~d (or 366~h). A cross section of the calcite distribution at the final time is shown in Figure~\ref{fig:cs_single}. We see that most calcite has precipitated in and around the entry of the leakage path, with a maximum value of 0.0483. From Figure~\ref{fig:obj_single} we see that $ \obj(\ctrl) $ have gone from negative to positive, indicating an increase in calcite precipitation, along with a decrease in injection and no-flow periods, during the iterations.

\begin{figure*}[h!]
	\subfloat[\label{fig:co2_single}]{\includegraphics[width=0.33\textwidth,height=0.3\textwidth]{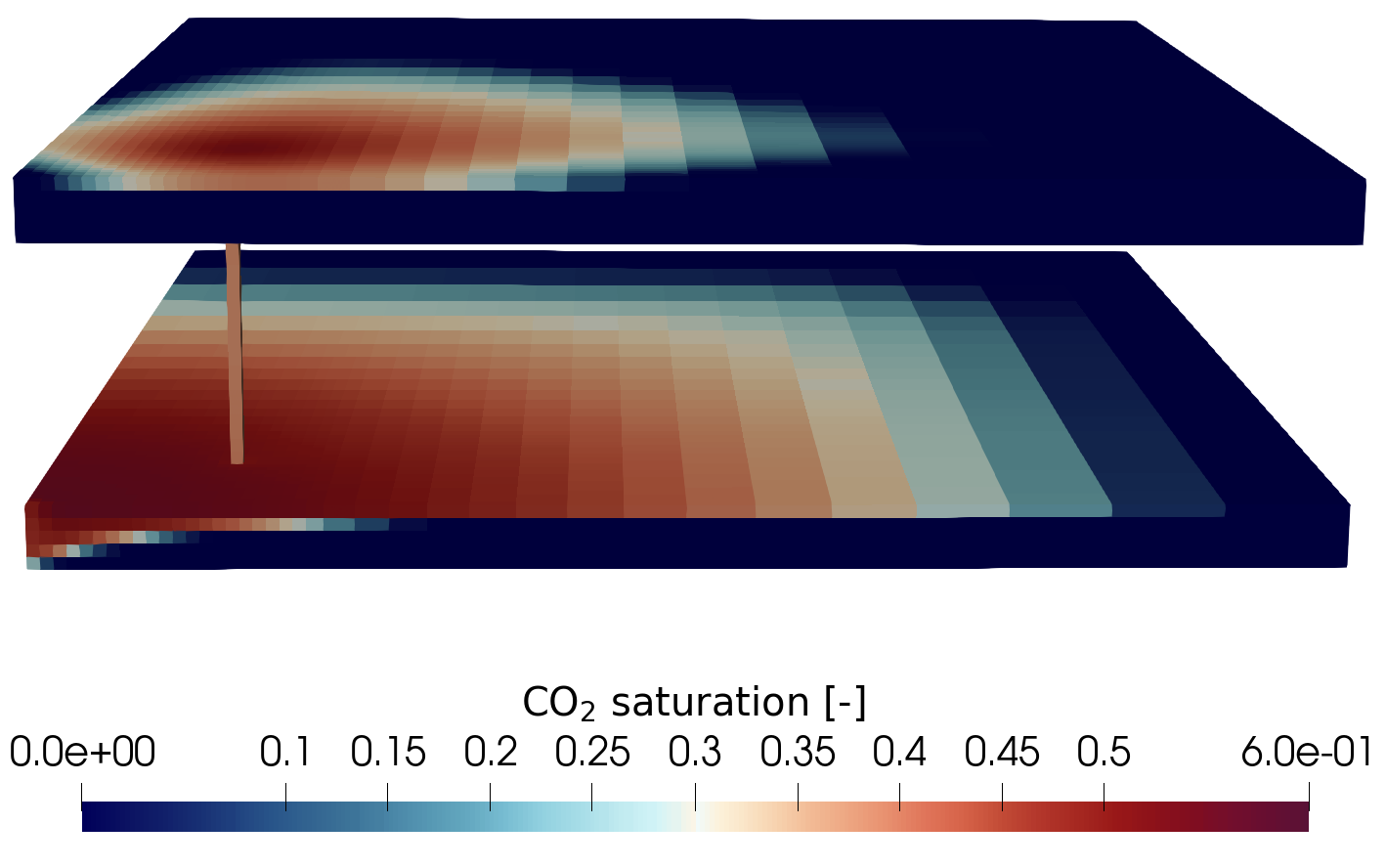}}
	\subfloat[\label{fig:co2micp_single}]{\includegraphics[width=0.33\textwidth,height=0.3\textwidth]{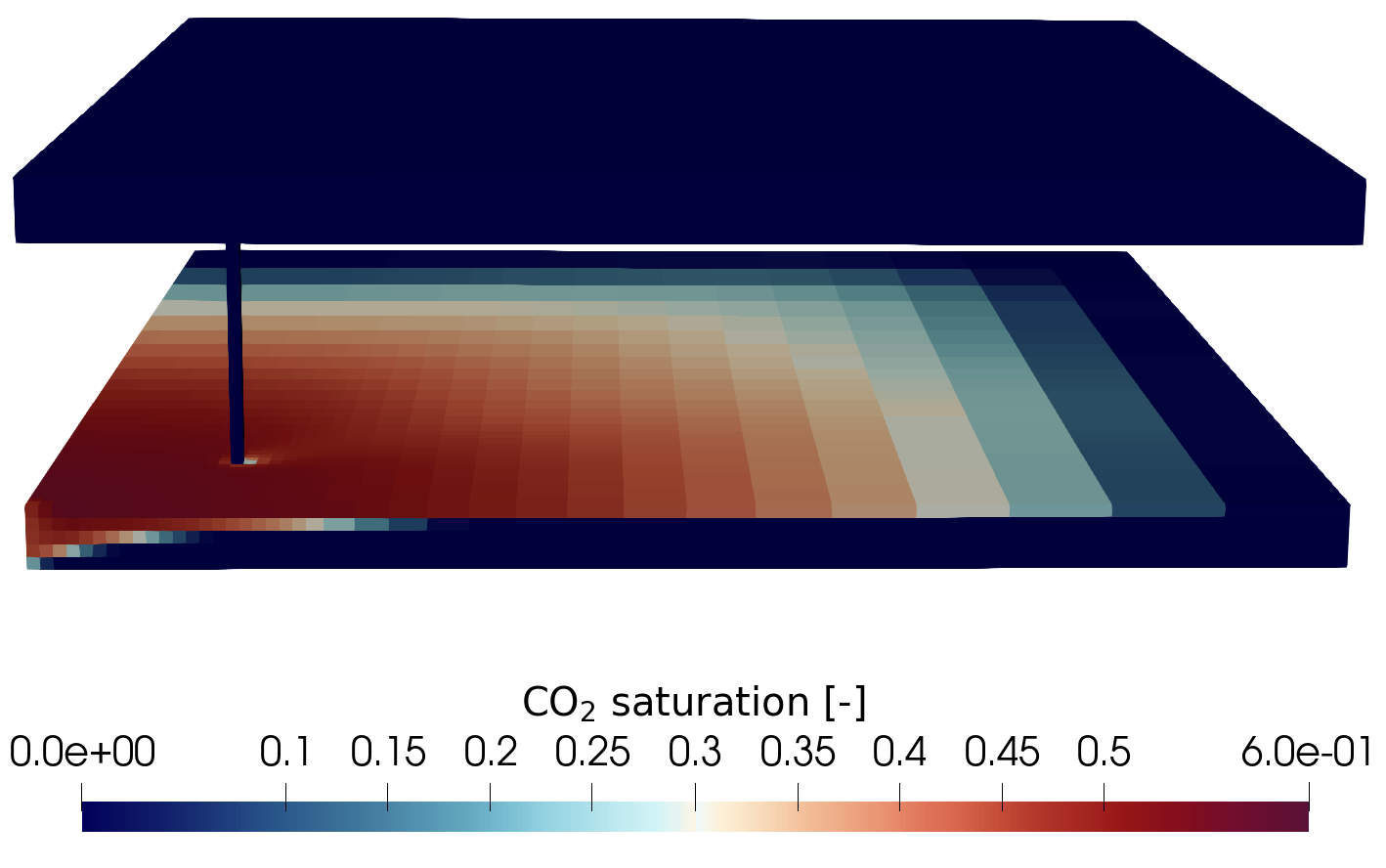}}
	\subfloat[\label{fig:leak_single}]{\includegraphics[width=0.33\textwidth]{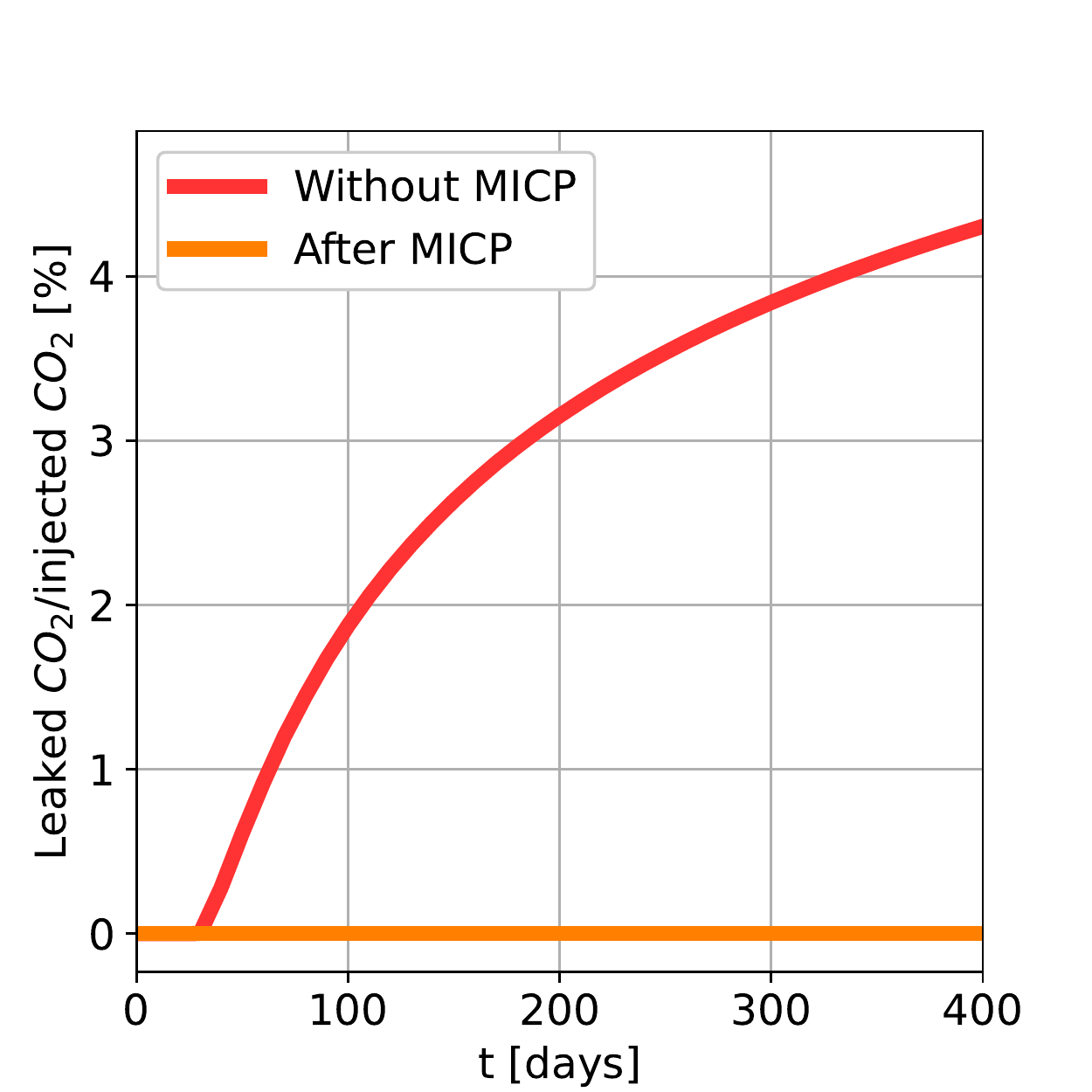}}
	\caption{Single leak: \co\ saturation after 400~d injection (a) before and (b) after MICP using the final optimization results, and (c) percentage of accumulated \co\ mass in upper aquifer over the injection period for both cases}\label{fig:co2inj_single}
\end{figure*}

\noindent Comparing Figure~\ref{fig:co2_single} and Figure~\ref{fig:co2micp_single}, we see the impact of the final calcite distribution on the \co\ injection simulation. From Figure~\ref{fig:leak_single} and the percentage reduction in \co\ mass in the upper aquifer in Table~\ref{tab:opt_results}, we see that the leakage path is essentially sealed.

\subsection{Double leak}\label{Scenario ii}
\begin{figure}[h!]
\centering
	\subfloat[\label{fig:cs_double}]{\includegraphics[width=0.33\textwidth]{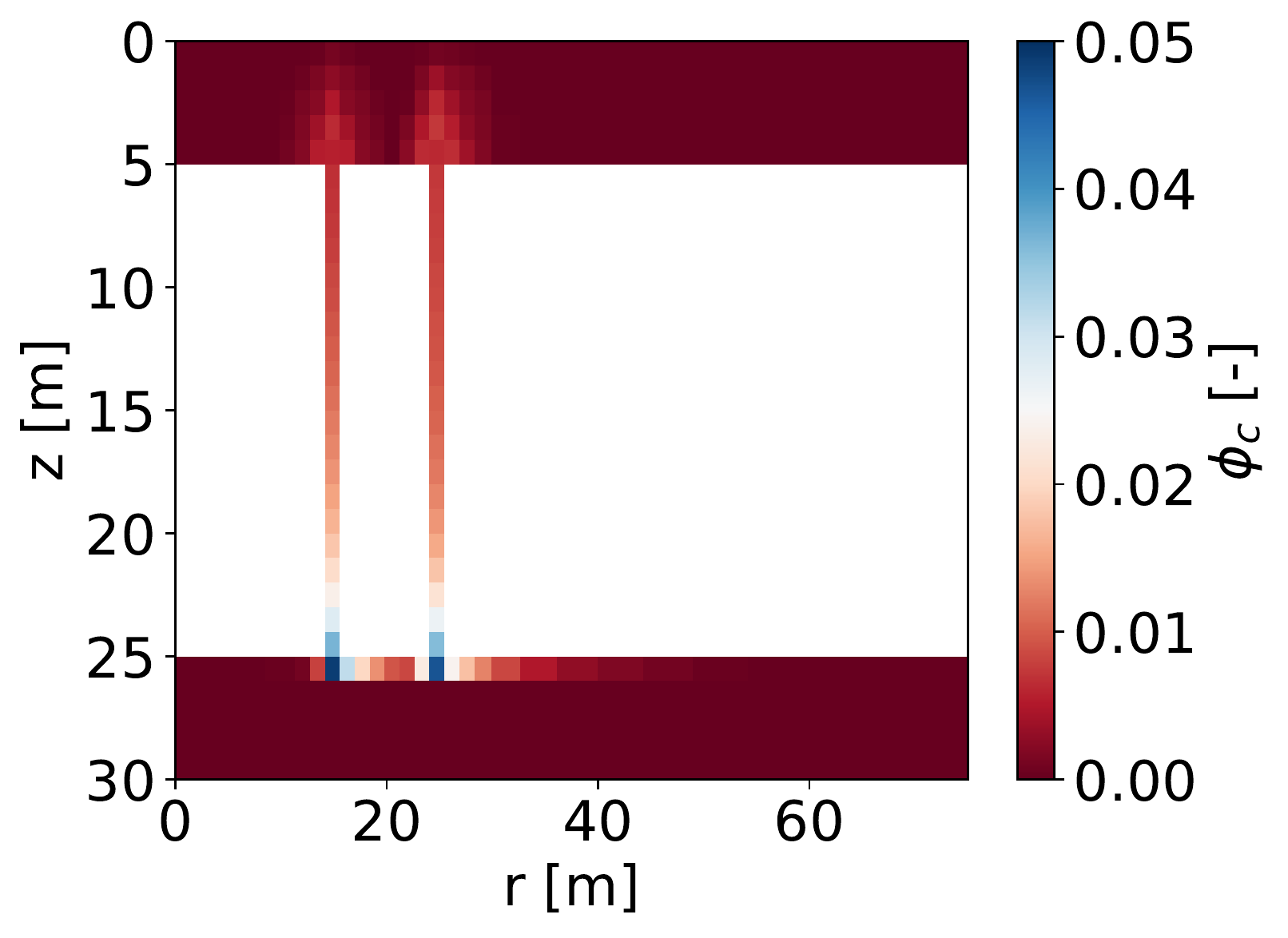}}\quad    
	\subfloat[\label{fig:obj_double}]{\includegraphics[width=0.335\textwidth]{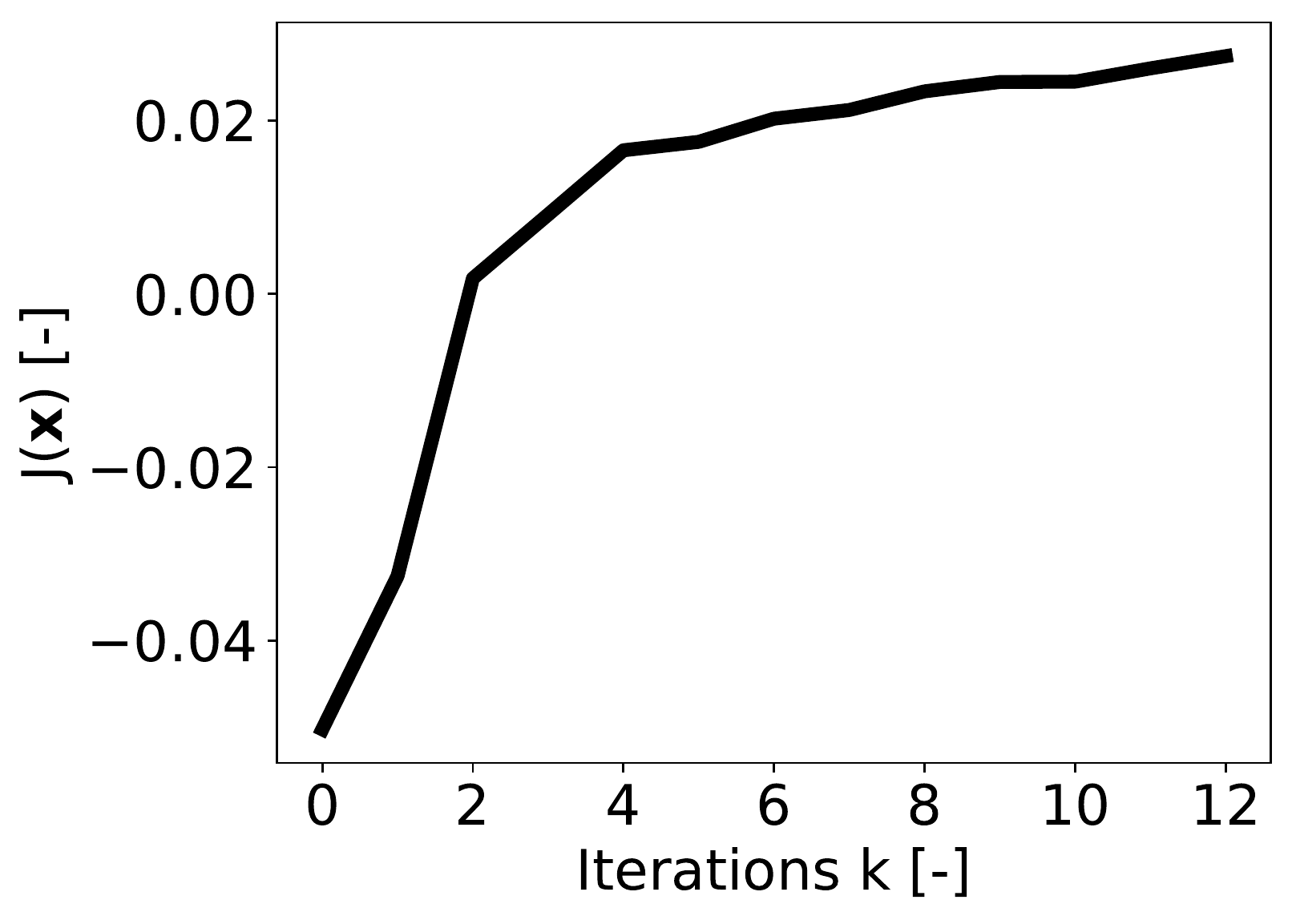}}
	\caption{Double leak: (a) Vertical cross-section plot from (x, y) = (0.0 m,1.0 m) to (x, y) = (90.5 m, 0.0 m) of the calcite
distribution using the final optimization result (the x-axis, r, indicates the cross-section length) and (b) objective function values versus iterations}
\end{figure}

In this scenario, the initial control variables were the same as for Single leak (cf. Section~\ref{Scenario i}), but $ \pen = 5.36\cdot10^{-3}$~d$ ^{-1} $. From Table~\ref{tab:opt_results}, we see that the final optimization time, $ t_3^{V} $, was 8.65~d (or 207.65~h). This is a significant reduction from the initial 15.25~d (or 366~h), but slightly higher than the final optimization time for Single leak. From the cross-section plot in Figure~\ref{fig:cs_double} we see that the calcite has precipitated in and around the entry of both leakage paths, with maximum value of 0.0487. Figure~\ref{fig:obj_double} shows again that $ \obj(\ctrl) $ have gone from negative to positive, thus calcite precipitation have increased along with decrease of total injection and no-flow time.

\begin{figure*}[h!]
	\subfloat[\label{fig:co2_double}]{\includegraphics[width=0.33\textwidth,height=0.3\textwidth]{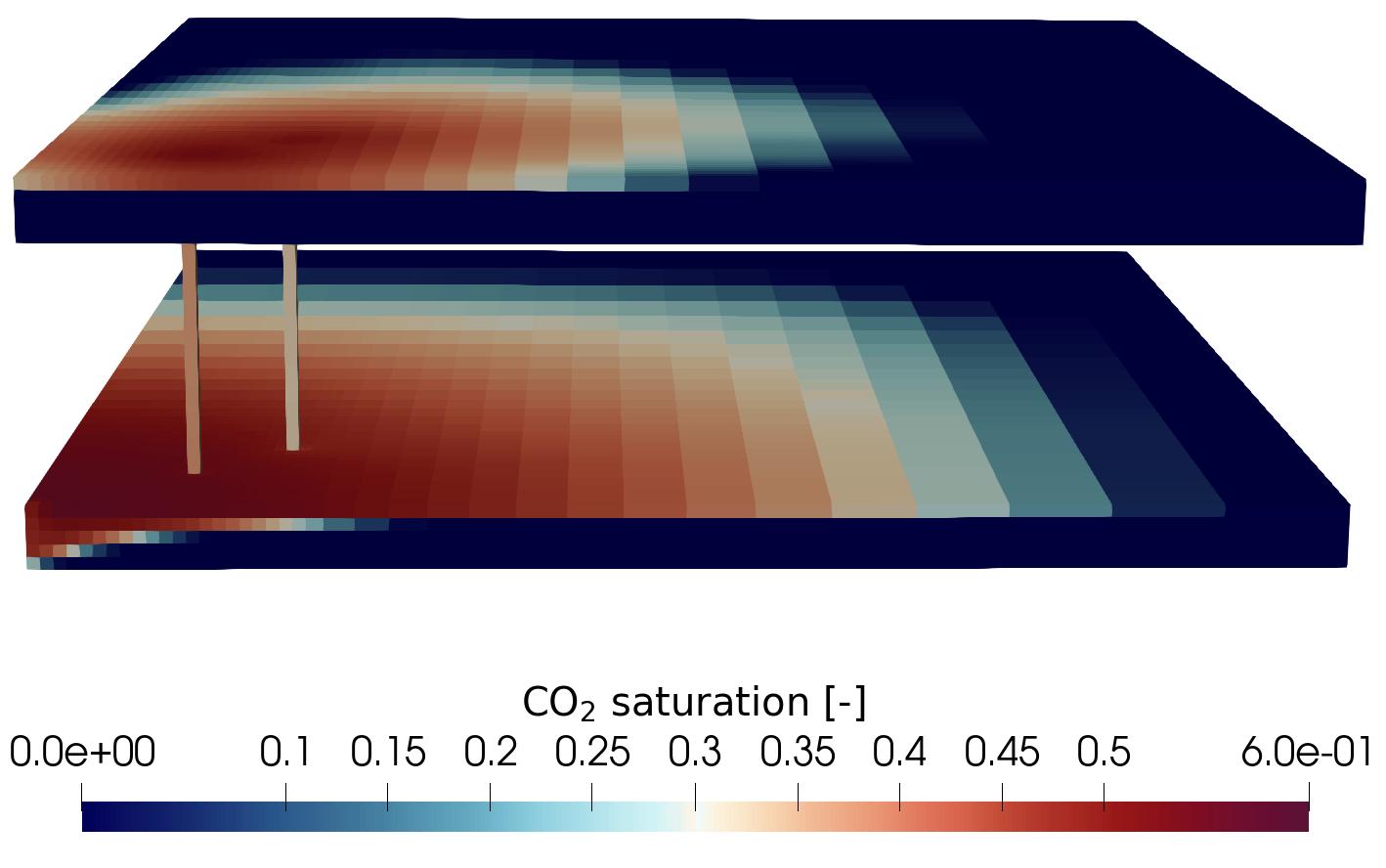}}
	\subfloat[\label{fig:co2micp_double}]{\includegraphics[width=0.33\textwidth,height=0.3\textwidth]{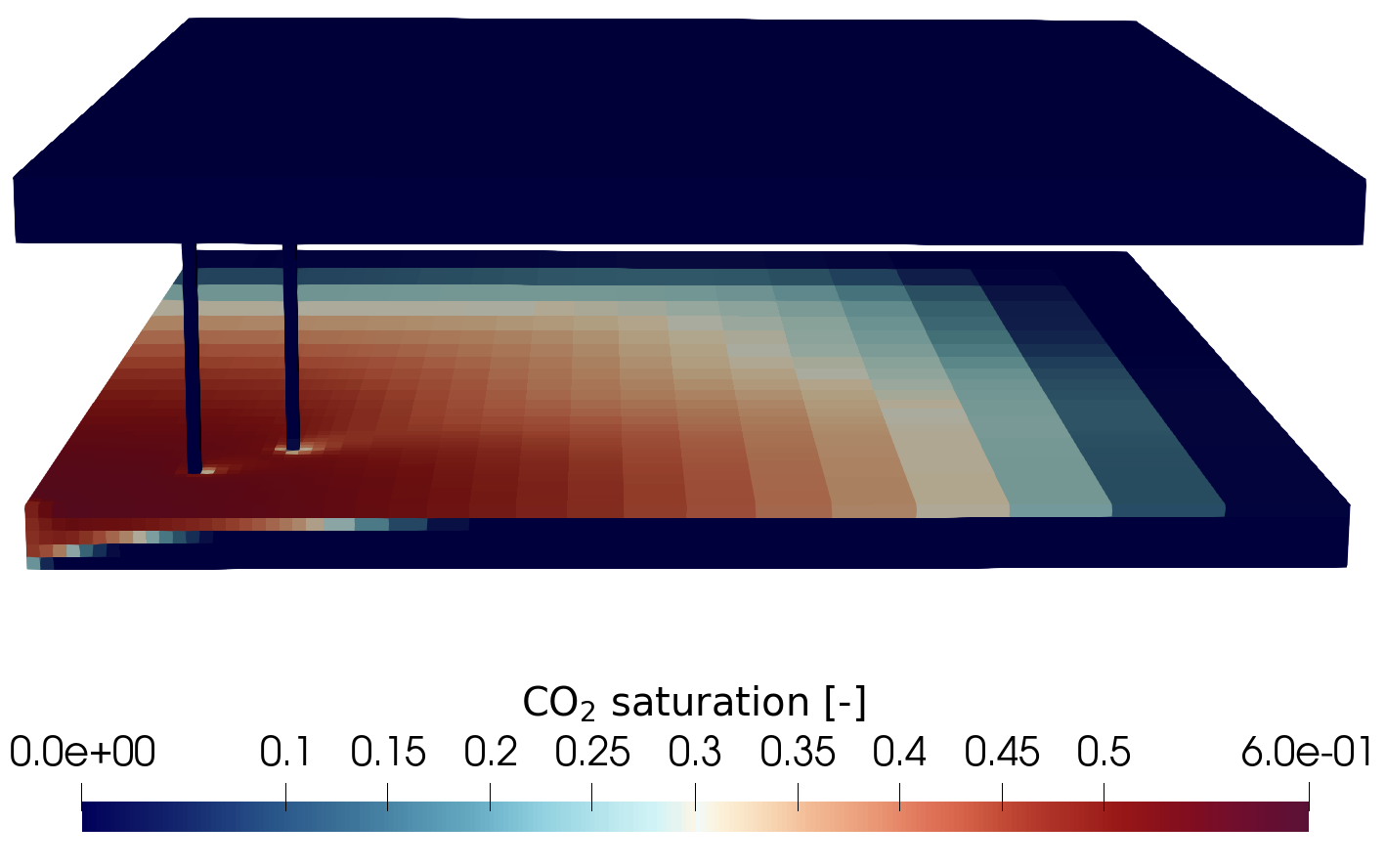}}
	\subfloat[\label{fig:leak_double}]{\includegraphics[width=0.33\textwidth]{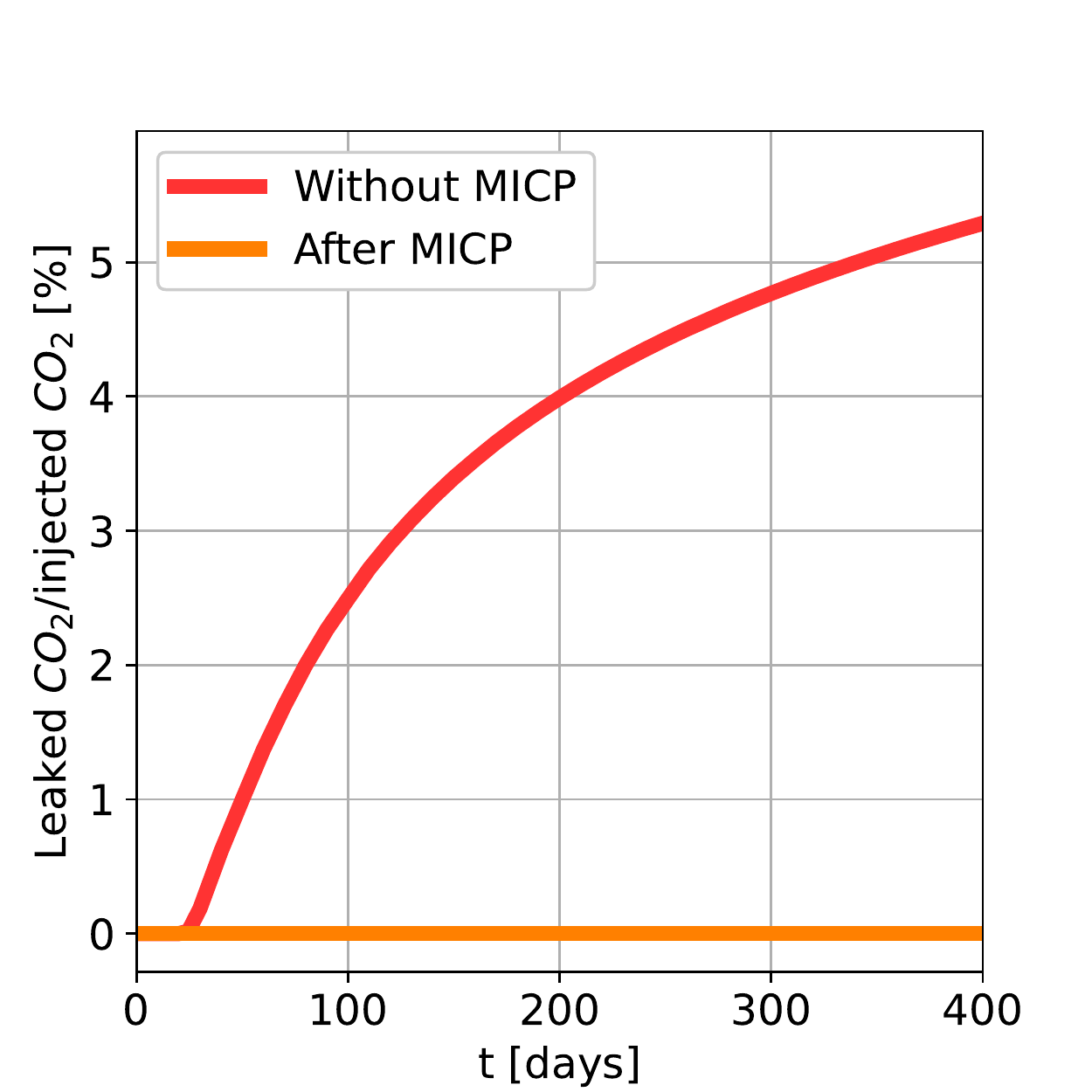}}
	\caption{Double leak: \co\ saturation after 400~d injection (a) before and (b) after MICP using the final optimization results, and (c) percentage of accumulated \co\ mass in upper aquifer over injection period for both cases}
\end{figure*}

\noindent  Figures~\ref{fig:co2_double} and~\ref{fig:co2micp_double} show that the \co\ leakage have been reduced significantly. From Figure~\ref{fig:leak_double} and the percentage reduction in \co\ mass in the upper aquifer in Table~\ref{tab:opt_results}, we see that the leakage path is essentially sealed.

\subsection{Diagonal leak}\label{Scenario iii}
\begin{figure}[h!]
\centering
	\subfloat[\label{fig:cs_diag}]{\includegraphics[width=0.33\textwidth]{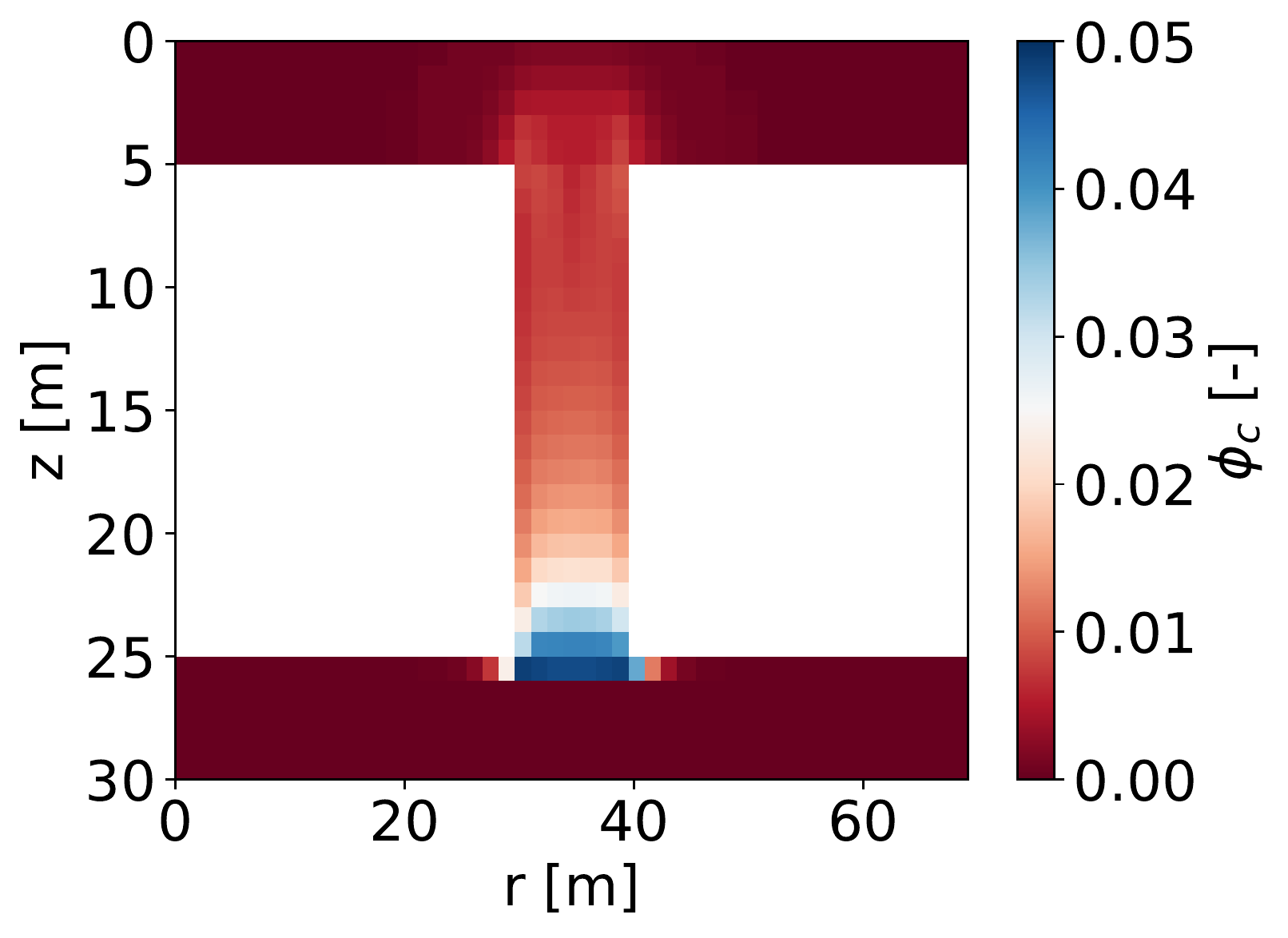}}\quad    
	\subfloat[\label{fig:obj_diag}]{\includegraphics[width=0.335\textwidth]{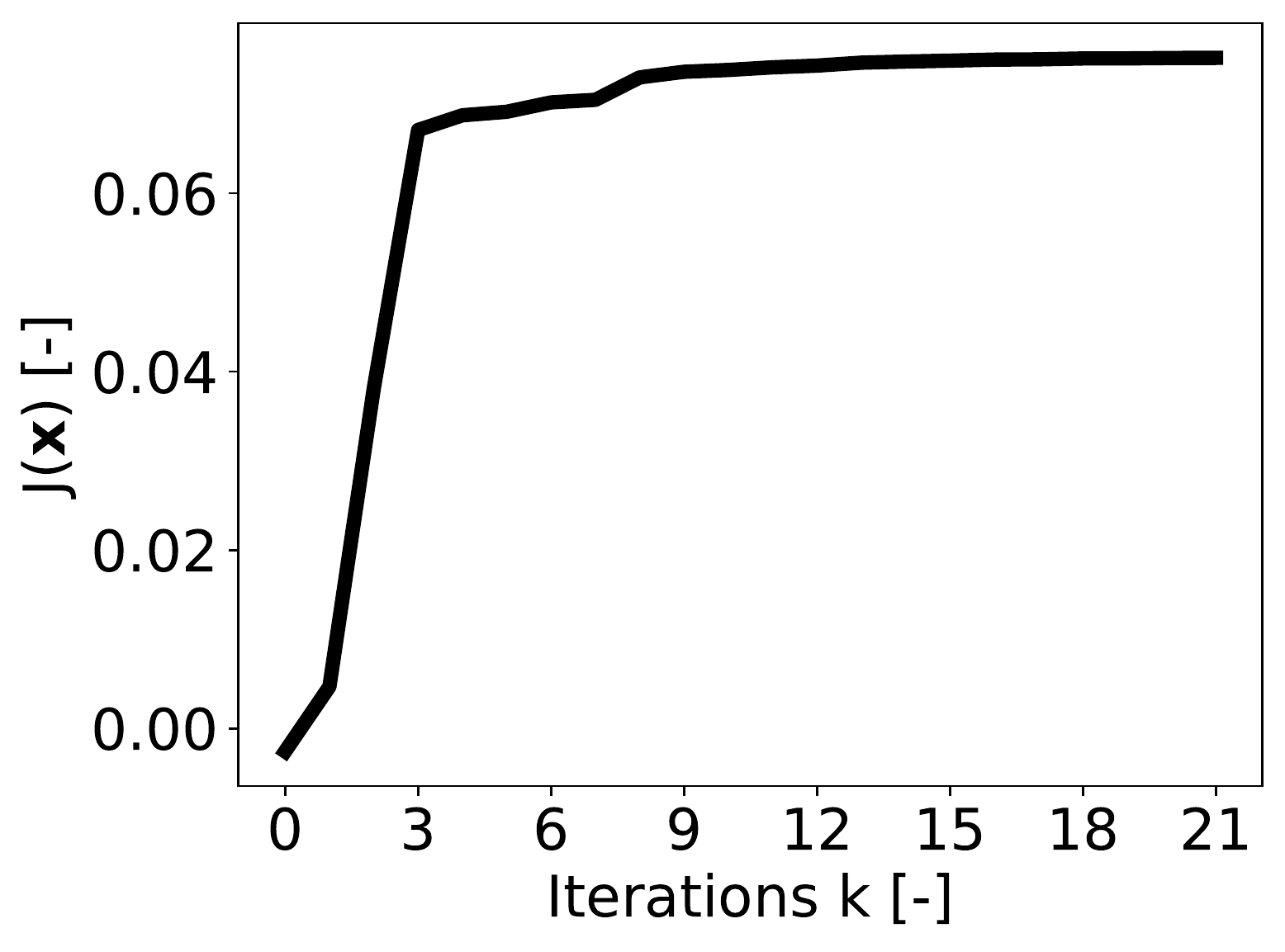}}
	\caption{Diagonal leak: (a) Vertical cross-section plot from (x, y) = (0.0 m, 1.0 m) to (x, y) = (90.5 m, 0.0 m) of the
calcite distribution using the final optimization result (the x-axis, r, indicates the cross-section length) and (b) objective function values versus iterations}
\end{figure}

In this scenario, the initial control variables were changed from the Single and Double leak scenarios, to 0.5~d for $ \conctrlcomp_0 $, 0.05~d for $ \conctrlwat_0 $, and 1~d for $ \conctrlnflw_0 $, in addition to $ \pen =  1.79\cdot10{-3}$~d$ ^{-1} $. Hence, we start with a short total time of 7.75~d (or 186~h). We compensate this by having a low $ \pen $, to put even more weight on the first term in \eqref{eq:obj} compared to the Single and Double leak scenarios. From Table~\ref{tab:opt_results} we see that the final optimization time is 8.31~d (or 199.5~h), which is not a significant increase from the total time of the initial control variables. Furthermore, the cross-section plot in Figure~\ref{fig:cs_diag} show that the calcite distribution in and around the entry of the leakage path is high, with a maximum value of 0.0484. Thus, the optimization have increased the calcite precipitation in the leakage path while keeping the total injection and no-flow time low, which is also what the objective function plot in Figure~\ref{fig:obj_diag} indicates.

\begin{figure*}[h!]
	\subfloat[\label{fig:co2_diag}]{\includegraphics[width=0.33\textwidth,height=0.3\textwidth]{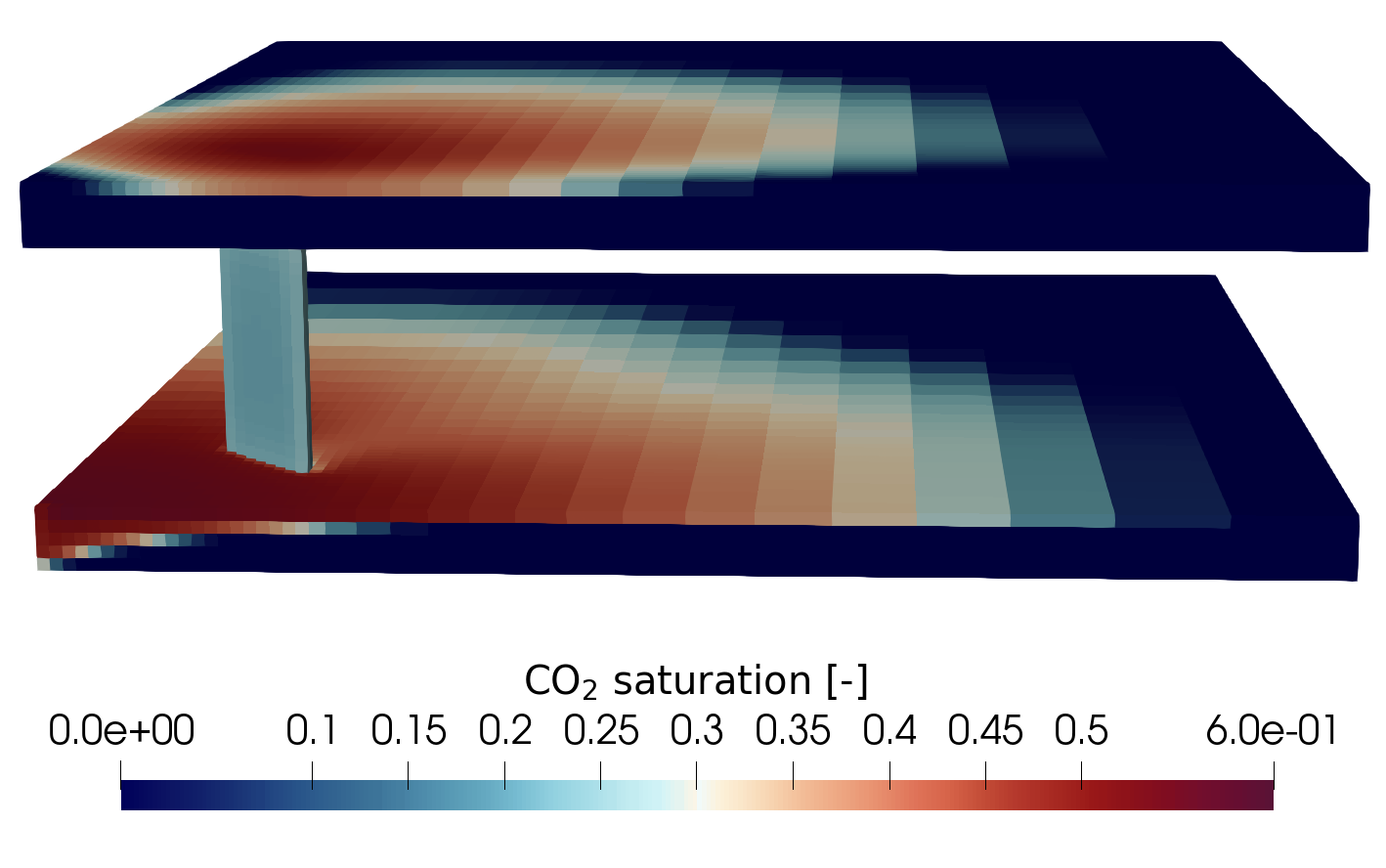}}
	\subfloat[\label{fig:co2micp_diag}]{\includegraphics[width=0.33\textwidth,height=0.3\textwidth]{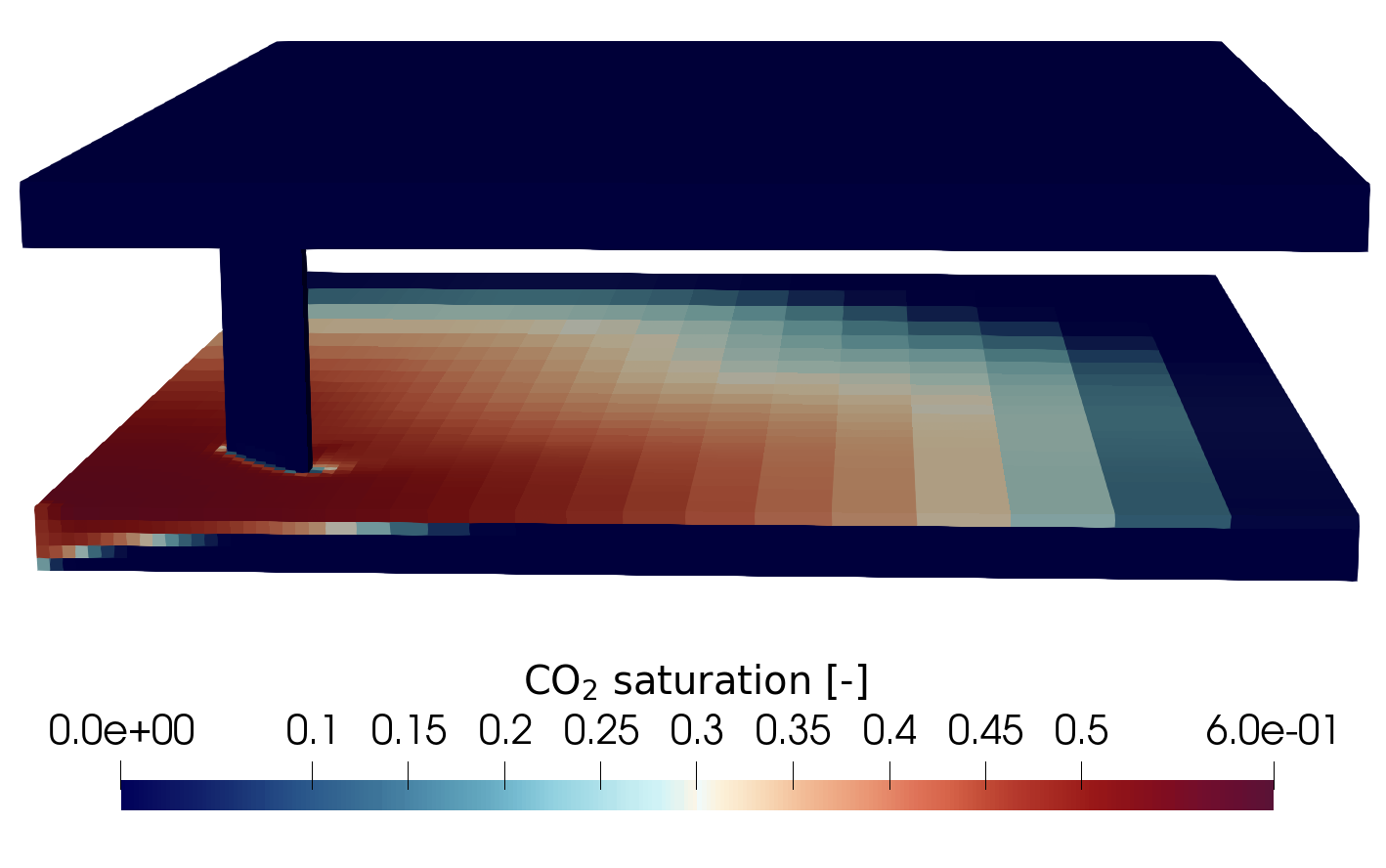}}
	\subfloat[\label{fig:leak_diag}]{\includegraphics[width=0.33\textwidth]{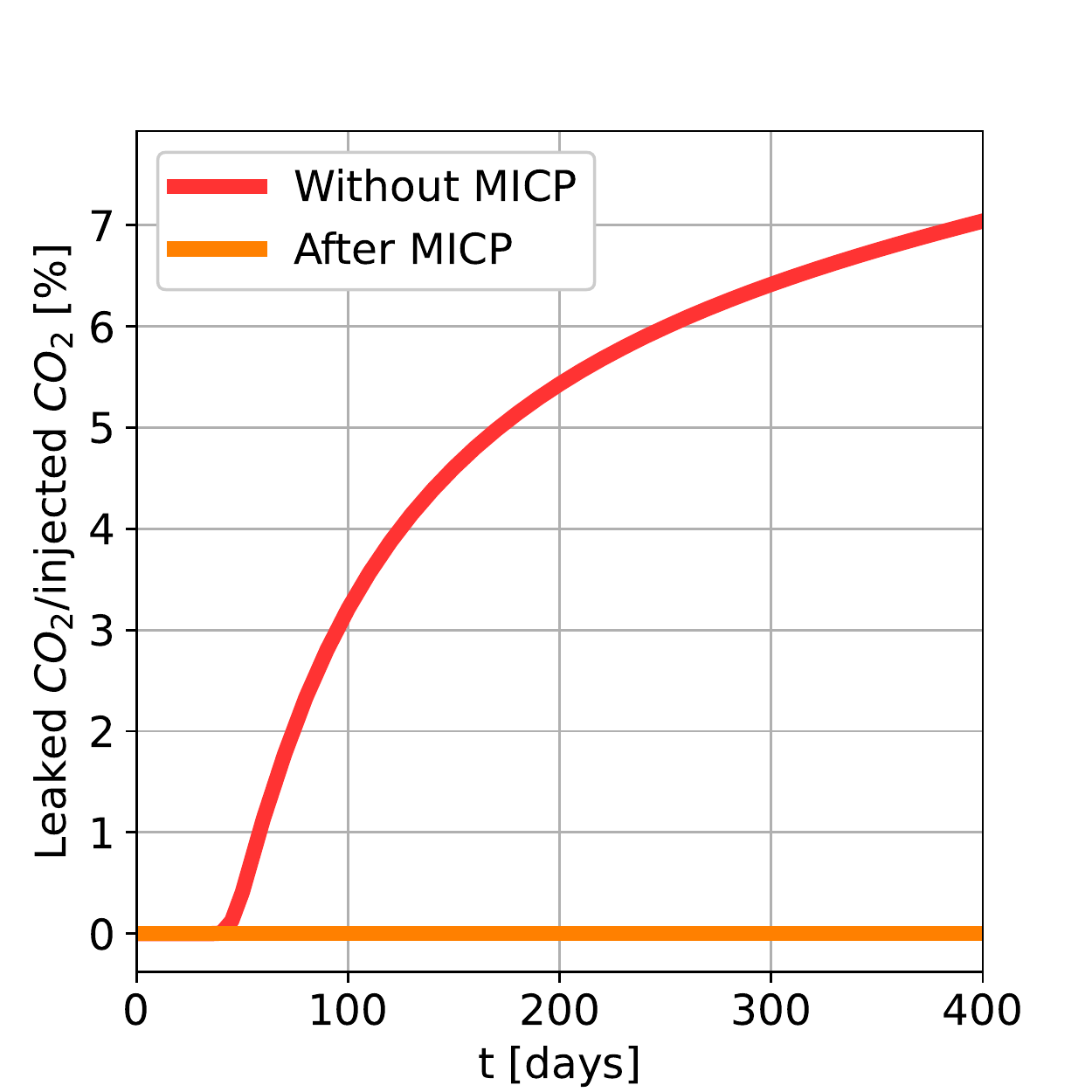}}
	\caption{Diagonal leak: \co\ saturation after 400~d injection (a) before and (b) after MICP using the final optimization results, and (c) percentage of accumulated \co\ mass in upper aquifer over injection period for both cases}\end{figure*}

\noindent In Figures~\ref{fig:co2_diag} and~\ref{fig:co2micp_diag} we see that the \co\ in the upper aquifer have been reduced significantly. From Figure~\ref{fig:leak_diag} and the percentage reduction in \co\ mass in the upper aquifer in Table~\ref{tab:opt_results}, we see that the leakage path is essentially sealed.

\section{Discussion}\label{Discussion}
From Table~\ref{tab:opt_results} we see that the injection period with the combined growth and cementation solutions in phase $ IV $ (i.e. $\dt_{1,3}^{IV} $) is the longest in all three scenarios. This indicates that when the biofilm has successfully been established in phases $ I $--$ III $, a long-time injection of growth and cementation solutions is needed for maximizing the sealing of the leakage paths. Furthermore, we see that longer no-flow periods are needed in phases $ I $--$ III $ compared to phases $ IV $ and $ V $. This confirms the observation that establishing a biofilm in phases $ I $--$ III $ is important. Lastly, we see that for the Single and Diagonal leak scenarios phase $ V $ is short, indicating four phases may be enough for efficient sealing, while a reasonable amount of time was spent in Double leak phase $ V $, thus indicating the need for five phases.

Simulating radial injection from a vertical well necessarily means that a lot of the microbial, growth, and cementation solutions are wasted during the injections, which would be even more prominent if we had simulated the full domain. Even though it might be possible from an engineering perspective to direct well flow in a specific direction in the reservoir, it is complex from a simulation perspective. A preliminary study simulating directional radial wells can be found in \texttt{py-micp}. When simulation tools for directional injection have been developed, economic terms for efficient use of the three injected solutions may be included in the objective function, such as upper limits on injected mass of expensive components.

\section{Conclusions}\label{Conclusions}
In this paper, we have presented and applied an optimization procedure for sealing leakage paths with MICP. The field-scale mathematical model for MICP processes were simulated using the industry-standard simulator OPM Flow, while the gradient-based EnOpt algorithm was used for the optimization. An injection strategy for field-scale application of MICP was developed, where growth and cementation solutions were injected at the same time but in different well segments, to efficiently initiate the MICP process at the leakage paths. Furthermore, we defined the objective function such that maximizing calcite precipitation at the leakage paths are weighted with total injection and no-flow time during the optimization. Hence, the economic aspect of having the shortest possible MICP operational time is considered in the optimizations.

The optimization procedure was applied to three synthetic \co\ leakage scenarios. The numerical results showed that the optimization procedure was able to essentially close the leakage path in the Single leak scenario, and almost completely seal the two leakage paths and wide, diagonal leakage path in the Double and Diagonal leak scenarios, respectively. In the optimizations, the total injection and no-flow time was low while the calcite distribution in and around the entry of the leakage paths were maximized. Hence, including economic aspects, like total operational time, in the objective function is recommended for MICP, and similar leakage sealing, optimization procedures.

\paragraph{Acknowledgements:} The authors are grateful for the financial support from Research Council of Norway through the project Efficient models for microbially induced calcite precipitation as a seal for \co\ storage (MICAP) (grant 268390). Open Access funding provided by NORCE Norwegian Research Centre AS. The authors thank Tor Harald Sandve for useful discussions and guidance for the implementation of the MICP model in OPM Flow.

\paragraph{Data availability:} OPM Flow, the numerical simulator with MICP functionality used in this study, can be obtained at \url{https://github.com/OPM/opm-simulators}. The MATLAB/GNU Octave code MRST, which also includes the MICP module \texttt{ad-micp}, was used to generate the grids and can be obtained at \url{https://www.sintef.no/projectweb/mrst/}. The Python package \texttt{py-micp} can be found at \url{https://github.com/daavid00/py-micp.git}.


\begin{thebibliography}{}

\bibitem[\protect\citeauthoryear{Baxendale et al.}{2021}]{Baxendale2021}
Baxendale, D. and et al.,
2021.
OPM Flow documentation manual.
\href{https://opm-project.org/?page\_id=955}{https://opm-project.org/?page\_id=955}.

\bibitem[\protect\citeauthoryear{Burbank et al.}{2011}]{Burbank2011}
Burbank, M.B., Weaver, T.J., Green, T.L., Williams, B., Crawford, R.L.,
2011.
Precipitation of calcite by indigenous microorganisms to strengthen liquefiable soils.
\textit{Geomicrobiol. J.}
\textbf{28}~(4),
301--312.
\href{https://doi.org/10.1080/01490451.2010.499929}{https://doi.org/10.1080/01490451.2010.499929}.

\bibitem[\protect\citeauthoryear{Chang et al.}{2020}]{Chang2020}
Chang, Y., Lorentzen, R.J., Nævdal, G., Feng, T.,
2020.
{OLYMPUS} optimization under geological uncertainty.
\textit{Computat. Geosci.}
\textbf{24}~(6),
2027--2042.
\href{https://doi.org/10.1007/s10596-019-09892-x}{https://doi.org/10.1007/s10596-019-09892-x}.

\bibitem[\protect\citeauthoryear{Chen et al.}{2009}]{Chen2009}
Chen, Y., Oliver, D.S., Zhang, D.,
2009.
Efficient ensemble-based closed-loop production optimization.
\textit{SPE J.}
\textbf{14}~(4),
634--645.
\href{https://doi.org/10.2118/112873-PA}{https://doi.org/10.2118/112873-PA}.

\bibitem[\protect\citeauthoryear{Chen and Voskov}{2020}]{Chen2020}
Chen, Y., Voskov, D.,
2020.
Optimization of {CO$_2$} injection using multi-scale reconstruction of composition transport.
\textit{Computat. Geosci.}
\textbf{24}~(2),
819--835.
\href{https://doi.org/10.1007/s10596-019-09841-8}{https://doi.org/10.1007/s10596-019-09841-8}.

\bibitem[\protect\citeauthoryear{Class et al.}{2009}]{Class2009}
Class, H., Ebigbo, A., Helmig, R., Dahle, H.K., Nordbotten, J.M., Celia, M.A., Audigane, P., Darcis, M., Ennis-King, J., Fan, Y., et al.,
2009.
A benchmark study on problems related to {CO$_2$} storage in geologic formations.
\textit{Computat. Geosci.}
\textbf{13}~(4),
409--434.
\href{https://doi.org/10.1007/s10596-009-9146-x}{https://doi.org/10.1007/s10596-009-9146-x}.

\bibitem[\protect\citeauthoryear{Cunningham et al.}{2019}]{Cunningham2019}
Cunningham, A.B., Class, H., Ebigbo, A., Gerlach, R., Phillips, A.J., Hommel, J.,
2019.
Field-scale modeling of microbially induced calcite precipitation.
\textit{Comput. Geosci.}
\textbf{23}~(2),
399--414.
\href{https://doi.org/10.1007/s10596-018-9797-6}{https://doi.org/10.1007/s10596-018-9797-6}.

\bibitem[\protect\citeauthoryear{Cunningham et al.}{2011}]{Cunningham2011}
Cunningham, A.B., Gerlach, R., Spangler, L., Mitchell, A.C., Parks, S., Phillips, A.,
2011.
Reducing the risk of well bore leakage of {CO$_2$} using engineered biomineralization barriers.
\textit{Energy Proced.}
\textbf{4},
5178--5185.
\href{https://doi.org/10.1016/j.egypro.2011.02.495}{https://doi.org/10.1016/j.egypro.2011.02.495}.

\bibitem[\protect\citeauthoryear{De Muynck et al.}{2010}]{DeMuynck2010}
De Muynck, W., De Belie, N., Verstraete, W.,
2010.
Microbial carbonate precipitation in construction materials: {A} review.
\textit{Ecol. Eng.}
\textbf{36}~(2),
118--136.
\href{https://doi.org/10.1016/j.ecoleng.2009.02.006}{https://doi.org/10.1016/j.ecoleng.2009.02.006}.

\bibitem[\protect\citeauthoryear{Do and Reynolds}{2013}]{Do2013}
Do, S.T., Reynolds, A.C.,
2013.
Theoretical connections between optimization algorithms based on an approximate gradient.
\textit{Computat. Geosci.}
\textbf{17}~(6),
959--973.
\href{https://doi.org/10.1007/s10596-013-9368-9}{https://doi.org/10.1007/s10596-013-9368-9}.

\bibitem[\protect\citeauthoryear{Ebigbo et al.}{2007}]{Ebigbo2007}
Ebigbo, A., Class, H., Helmig, R.,
2007.
{CO$_2$} leakage through an abandoned well: problem-oriented benchmarks.
\textit{Computat. Geosci.}
\textbf{11}~(2),
103--115.
\href{https://doi.org/10.1007/s10596-006-9033-7}{https://doi.org/10.1007/s10596-006-9033-7}.

\bibitem[\protect\citeauthoryear{Ebigbo et al.}{2012}]{Ebigbo2012}
Ebigbo, A., Phillips, A., Gerlach, R., Helmig, R., Cunningham, A.B., Class, H., Spangler, L.H.,
2012.
Darcy-scale modeling of microbially induced carbonate mineral precipitation in sand columns.
\textit{Water Resour. Res.}
\textbf{48}~(7),
W07519.
\href{https://doi.org/10.1029/2011WR011714}{https://doi.org/10.1029/2011WR011714}.

\bibitem[\protect\citeauthoryear{Elenius et al.}{2018}]{Elenius2018}
Elenius, M., Skurtveit, E., Yarushina, V., Baig, I., Sundal, A., Wangen, M., Land- schulze, K., Kaufmann, R., Choi, J.C., Hell- evang, H., et al.,
2018.
Assessment of {CO$_2$} storage capacity based on sparse data: {Skade Formation}.
\textit{Int. J. Greenh. Gas Con.}
\textbf{79},
252--271.
\href{https://doi.org/10.1016/j.ijggc.2018.09.004}{https://doi.org/10.1016/j.ijggc.2018.09.004}.

\bibitem[\protect\citeauthoryear{Haszeldine et al.}{2018}]{Haszeldine2018}
Haszeldine, R.S., Flude, S., Johnson, G., Scott, V.,
2018.
Negative emissions technologies and carbon capture and storage to achieve the Paris Agreement commitments.
\textit{Philos. T. Roy. Soct. A.}
\textbf{376},
20160447.
\href{https://doi.org/10.1098/rsta.2016.0447}{https://doi.org/10.1098/rsta.2016.0447}.

\bibitem[\protect\citeauthoryear{Hodneland et al.}{2019}]{Hodneland2019}
Hodneland, E., Gasda, S., Kaufmann, R., Bekkvik, T.C., Hermanrud, C., Midttømme, K.,
2019.
Effect of temperature and concentration of impurities in the fluid stream on {CO$_2$} migration in the {Utsira} formation.
\textit{Int. J. Greenh. Gas Con.},
\textbf{83},
20--28.
\href{https://doi.org/10.1016/j.ijggc.2019.01.020}{https://doi.org/10.1016/j.ijggc.2019.01.020}.

\bibitem[\protect\citeauthoryear{Hommel et al.}{2015}]{Hommel2015}
Hommel, J., Lauchnor, E., Phillips, A., Gerlach, R., Cunningham, A.B., Helmig, R., Ebigbo, A., Class, H.,
2015.
A revised model for microbially induced calcite precipitation: {I}mprovements and new insights based on recent experiments.
\textit{Water Resour. Res.}
\textbf{51}~(5),
3695--3715.
\href{https://doi.org/10.1002/2014WR016503}{https://doi.org/10.1002/2014WR016503}.

\bibitem[\protect\citeauthoryear{Landa-Marb\'an et al.}{2021b}]{Landa-Marban2021b}
Landa-Marb\'an, D., Kumar, K., Tveit, S., Gasda, S.E.,
2021.
Numerical studies of {CO$_2$} leakage remediation by {M}{I}{C}{P}-based plugging technology.
\textit{SINTEF Academic Press}.
284--290.
\href{https://hdl.handle.net/11250/2786420}{https://hdl.handle.net/11250/2786420}.

\bibitem[\protect\citeauthoryear{Landa-Marb\'an et al.}{2021a}]{Landa-Marban2021}
Landa-Marb\'an, D., Tveit, S., Kumar, K., Gasda, S.E.,
2021.
Practical approaches to study microbially induced calcite precipitation at the field scale.
\textit{Int. {J}. {G}reenh. {G}as {C}ont.}
\textbf{106},
103256.
\href{https://doi.org/10.1016/j.ijggc.2021.103256}{https://doi.org/10.1016/j.ijggc.2021.103256}.

\bibitem[\protect\citeauthoryear{Minto et al.}{2019}]{Minto2019}
Minto, J.M., Lunn, R.J., El Mountassir, G.,
2019.
Development of a reactive transport model for field-Scale simulation of microbially induced carbonate precipitation.
\textit{Water Resour. Res.}
\textbf{55}~(8),
7229--7245.
\href{https://doi.org/10.1029/2019WR025153}{https://doi.org/10.1029/2019WR025153}.

\bibitem[\protect\citeauthoryear{Mulrooney et al.}{2020}]{Mulrooney2020}
Mulrooney, M.J., Osmond, J.L., Skurtveit, E., Faleide, J.I., Braathen, A.,
2020.
Structural analysis of the {Smeaheia} fault block, a potential {CO$_2$} storage site, northern {Horda Platform, North Sea}.
\textit{Mar. Petrol. Geol.}
\textbf{121},
104598.
\href{https://doi.org/10.1016/j.marpetgeo.2020.104598}{https://doi.org/10.1016/j.marpetgeo.2020.104598}.

\bibitem[\protect\citeauthoryear{Nassar et al.}{2018}]{Nassar2018}
Nassar, M.K., Gurung, D., Bastani, M., Ginn, T.R., Shafei, B., Gomez, M.G., Graddy, C.M.R., Nelson, D.C., DeJong, J.T.,
2018.
Large-scale experiments in microbially induced calcite precipitation ({MICP}): Reactive transport model development and prediction.
\textit{Water Resour. Res.},
\textbf{54},
480-500.
\href{https://doi.org/10.1002/2017WR021488}{https://doi.org/10.1002/2017WR021488}.

\bibitem[\protect\citeauthoryear{Phillips et al.}{2016}]{Phillips2016}
Phillips, A.J., Cunningham, A.B., Gerlach, R., Hiebert, R., Hwang, C., Lomans, B.P., Westrich, J., Mantilla, C., Kirksey, J., Esposito, R., Spangler, L.,
2016.
Fracture sealing with microbially-induced calcium carbonate precipitation: A field study.
\textit{Environ. Sci. Technol.}
\textbf{50}~(7),
4111--4117.
\href{https://doi.org/10.1021/acs.est.5b05559}{https://doi.org/10.1021/acs.est.5b05559}.

\bibitem[\protect\citeauthoryear{Phillips et al.}{2013a}]{Phillips2013}
Phillips, A.J., Gerlach, R., Lauchnor, E., Mitchell, A.C., Cunningham, A.B., Span- gler, L.,
2013.
Engineered applications of ureolytic biomineralization: {A} review.
\textit{Biofouling},
\textbf{29}~(6),
715--733.
\href{https://doi.org/10.1080/08927014.2013.796550}{https://doi.org/10.1080/08927014.2013.796550}.

\bibitem[\protect\citeauthoryear{Phillips et al.}{2013b}]{Phillips2013a}
Phillips, A.J., Lauchnor, E., Eldring, J., Esposito, R., Mitchell, A.C., Gerlach, R., Cunningham, A.B., Spangler, L.H.,
2013
Potential {CO$_2$} leakage reduction through biofilm-induced calcium carbonate precipitation.
\textit{Environ. Sci. Technol.}
\textbf{47},
142--149.
\href{https://doi.org/10.1021/es301294q}{https://doi.org/10.1021/es301294q}.

\bibitem[\protect\citeauthoryear{Rasmussen et al.}{2019}]{Rasmussen2019}
Rasmussen, A.F., Sandve, T.H., Bao, K., Lauser, A., Hove, J., Skaflestad, B., Klofkorn, R., Blatt, M., Rustad, A.B., Savareid, O., Lie, K., Thune, A.,
2019.
The {Open Porous Media Flow} reservoir simulator.
\textit{Comput. {M}ath. {A}ppl.}
\textbf{81},
159--185.
\href{https://doi.org/10.1016/j.camwa.2020.05.014}{https://doi.org/10.1016/j.camwa.2020.05.014}.

\bibitem[\protect\citeauthoryear{Stordal et al.}{2016}]{Stordal2016}
Stordal, A.S., Szklarz, S.P., Leeuwenburgh, O.,
2016.
A theoretical look at ensemble-based optimization in reservoir management.
\textit{Math. Geosci.}
\textbf{48}~(4),
399--417.
\href{https://doi.org/10.1007/s11004-015-9598-6}{https://doi.org/10.1007/s11004-015-9598-6}.

\bibitem[\protect\citeauthoryear{Tveit et al.}{2018}]{Tveit2018}
Tveit, S., Gasda, S.E., Hægland, H., B{\o}dtker, G., Elenius, M.,
2018.
Numerical study of microbially induced calcite precipitation as a leakage mitigation solution for {CO$_2$} storage.
\textit{European Association of Geoscientists \& Engineers}.
\href{https://doi.org/10.3997/2214-4609.201802956}{https://doi.org/10.3997/2214-4609.201802956}.

\bibitem[\protect\citeauthoryear{Tveit et al.}{2020}]{Tveit2020}
Tveit, S., Pettersson, P., Landa-Marb\'an, D.,
2020.
Optimizing sealing of {CO$_2$} leakage paths with microbially induced calcite precipitation under uncertainty.
\textit{European Association of Geoscientists \& Engineers}.
\href{https://doi.org/10.3997/2214-4609.202035087}{https://doi.org/10.3997/2214-4609.202035087}.

\bibitem[\protect\citeauthoryear{Wu et al.}{2017}]{Wu2017}
Wu, J., Wang, X.-B., Wang, H.-F., Zeng, R.J.,
2017.
Microbially induced calcium carbonate precipitation driven by ureolysis to enhance oil recovery.
\textit{{RSC} Adv.}
\textbf{7}~(59),
37382--37391.
\href{https://doi.org/10.1039/c7ra05748b}{https://doi.org/10.1039/c7ra05748b}.

\bibitem[\protect\citeauthoryear{Zhang and Agarwal}{2012}]{Zhang2012}
Zhang, Z., Agarwal, R.K.,
2012.
Numerical simulation and optimization of {CO$_2$} sequestration in saline aquifers for vertical and horizontal well injection.
\textit{Computat. Geosci.}
\textbf{16}~(4),
891--899.
\href{https://doi.org/10.1007/s10596-012-9293-3}{https://doi.org/10.1007/s10596-012-9293-3}. 

\end{thebibliography}
\end{document}